\documentclass[aps,pre,reprint,10pt,showpacs,showkeys,amsmath,amssymb]{revtex4-1}

\usepackage{amsfonts}
\usepackage{graphicx}
\usepackage{dcolumn}
\usepackage{bm} 
\usepackage{multirow}
\usepackage{topcapt}

\begin{document}
%
%

%
%
\title{Localization Properties of Covariant Lyapunov Vectors} 
\author{G. P. Morriss}
\email{g.morriss@unsw.edu.au}
\affiliation{School of Physics, University of New South Wales, Sydney NSW 2052, Australia}
\affiliation{Service de Physique de l'\'Etat Condens\'e, CEA - Saclay, 91191 Gif-sur-Yvette, France}

\date{\today}
\begin{abstract}
   The Lyapunov exponent spectrum and covariant Lyapunov vectors are studied for a quasi-one-dimensional system of hard disks as a function of density and system size. We characterize the system using the angle distributions between covariant vectors and the localization properties of both Gram-Schmidt and covariant vectors. At low density there is a {\it kinetic regime} that has simple scaling properties for the Lyapunov exponents and the average localization for part of the spectrum. This regime shows strong localization in a proportion of the first Gram-Schmidt and covariant vectors and this can be understood as highly localized configurations dominating the vector. The distribution of angles between neighbouring covariant vectors has characteristic shapes depending upon the difference in vector number, which vary over the continuous region of the spectrum. At dense gas or liquid like densities the behaviour of the covariant vectors are quite different. The possibility of tangencies between different components of the unstable manifold and between the stable and unstable manifolds is explored but it appears that exact tangencies do not occur for a generic chaotic trajectory. 
\end{abstract}
\maketitle  
%
\section{Introduction}

   The study of the Lyapunov vectors of particle systems generated by the Benettin scheme \cite{Ben,Shi79} - the so-called Gram-Schmidt (GS) vectors - is now well advanced for quasi-one-dimensional systems \cite{TMg} and two dimensional systems \cite{Posg}.
   The existence of a stepwise structure in the smallest non zero Lyapunov exponents is an extensive feature of the Lyapunov spectrum (the full set of Lyapunov exponents)  which has been well studied  \cite{Posg,TMg, For04,Yan05}. 
   For each step in the Lyapunov exponent spectrum the associated Lyapunov vector has a delocalised wavelike structure - referred to as a Lyapunov mode - which is of a particular type; either transverse, longitudinal or momentum proportional \cite{MT09}. 
   These modes are connected to the slowest macroscopic fluid properties (the dynamically conserved quantities) or equivalently the transport of mass, momentum or energy. 
   Transport of conserved quantities is always the slowest processes in particle systems thus they relate to the smallest instabilities or smallest Lyapunov exponents. 
   
   The statistical mechanics of chaotic many particle systems is an illustrative example of a probabilistic treatment of the global behavior of deterministic microscopic dynamics. 
   A great deal of effort has been devoted to finding links between macroscopic fluid quantities, such as transport coefficients, and chaotic properties of microscopic systems such as the Lyapunov exponents  \cite{Gas98,Dor99,EM08}.
   There have been some significant successes such as the conjugate pairing rule for the Lyapunov spectrum in some thermostated systems \cite{ECM90,DM96,TMth} and the fluctuation theorem \cite{ECM93,GC95} yet there have also been illustrations of non-chaotic systems with well defined transport properties.

   Microscopic particle systems that are models of real fluids can also be described from a macroscopic viewpoint.  
   The description begins with conservation equations and leads to the Navier-Stokes equations of fluid dynamics.
   The time evolution of the macroscopic quantities in the Navier-Stokes equations can also be studied as a chaotic dynamical system but their phase space behavior is fundamentally different to the phase space behavior of the original particle system.
   Nonlinear, dissipative, partial differential equations (PDEs) are frequently used to study natural phenomena in many areas of physics.
    Generic PDEs such as the Kuramoto-Sivashinsky equation and the complex Ginzberg-Landau equation have trajectories that are first exponentially attracted to a finite-dimensional inertial manifold and then settle in to a smooth global attractor.
    It is believed that a study of the covariant vectors or modes may lead to the identification of the inertial manifold \cite{TYGRC11}.
    The role played by the step structure in the Lyapunov spectrum for PDEs appears to be associated with convergence to the inertial manifold whereas for particle systems it is associated with the mechanically conserved quantities of the dynamics.
    Despite this difference in phase variables the angle distributions between covariant Lyapunov modes in both particle systems and PDEs seem to show some evidence of universal behaviour.

   The study of the covariant (C) Lyapunov vectors is less well understood but much progress has been made \cite{YT09, KK09, BP10, TYGRC11, TCGPT11}. 
   Covariant Lyapunov vectors \cite{TGC09} follow the physically important features of the nonlinear dynamics as they align with the instantaneous directions of the stable and unstable manifolds. 
   Thus instantaneous covariant vectors give information about the angles between various manifolds in the multidimensional phase space dynamics, either angles between two different stable manifolds, two different unstable manifolds, or between stable and unstable manifolds.
   
   In previous studies of the GS Lyapunov vectors an entropy based localization order parameter \cite{TM03b} was used as a probe of the structure of the vectors, and some aspects of the structure of the Gram-Schmidt vectors were obtained.
   Here we look more closely at the localization of the covariant vectors on average, the probability distribution of the localization and particular configurations which lead to features in both the distributions and instantaneous values of the order parameter.

    The manuscript is organized as follows. 
    An introduction to the method of calculating the covariant Lyapunov vectors. 
    The description of the QOD model and the GS results for the Lyapunov spectrum and the Lyapunov modes.
    Some mention of the scalar properties of the system.
    The bulk of the paper introduces the Localization order parameter and discusses in detail the results which follow; the average localization, distributions of localization, configurations which lead to strong localization, instantaneous localization, strong localization and the scaling of localization.
    Next we look in detail at the angle distributions between all covariant Lyapunov vectors in the system and discuss the possibilities of tangencies.
    We finish with a review of our semi-analytic knowledge of the behaviour of the covariant Lyapunov modes based on an understanding of the dynamics of the GS Lyapunov modes.


\section{Covariant Lyapunov Vectors}

   The Benettin scheme for calculating the Lyapunov spectrum consists of time evolving a set of orthogonal basis vectors $g^{(j)}_{n-1}$ which span the tangent space.
   Here $j$ is the number of the basis vector and $n-1$ is the step number (or discrete time).
   These vectors are periodically re-orthogonalized using the Gram-Schmidt procedure and the exponential growth and decay rates of each vector give the Lyapunov exponents.
   This set of orthogonal vectors are referred to as the Gram-Schmidt (GS) Lyapunov vectors.
   
   A number of methods for calculating covariant Lyapunov vectors have recently been proposed \cite{WS07,Sz07} but the method we follow here is that due to Ginelli et. al. \cite{GPTCLP07}, see also \cite{KK09} for more details.
   Making  the normalized Gram-Schmidt vector $g^{(j)}_{n-1}$, at phase point ${\bf x}_{n-1}$, the $j^{th}$ column of the matrix $G_{n-1}$, the forward time evolution in tangent space involves  two steps.
   First, evolve the system forward in time using the tangent space dynamics $J_{n-1}$ so that $\tilde{g}^{(j)}_{n} = J_{n-1} g^{(j)}_{n-1}$, and second, orthonormalize using a QR decomposition.
   So in matrix form the push forward in time is
\begin{equation}\label{g}
\tilde{G}_{n} = J_{n-1} G_{n-1}
\end{equation}   
and the QR decomposition is
\begin{equation}\label{qr}
\tilde{G}_{n} = G_{n} R_{n}.
\end{equation}  
where the columns of $G_{n}$ are orthogonal.
 
The $j^{th}$ covariant vector $v^{(j)}_{m}$ is contained in the sub-space spanned by the first $j$ GS vectors, so the matrix $V_{m}$ which has covariant vectors as its columns, can be written as
\begin{equation}\label{vm}
V_{m} = G_{m} C_{m}
\end{equation}   
where $C_{m}$ is an upper diagonal matrix of coefficients.
   The time evolution of $V_{m}$ is constructed from the time evolution of $G_{m}$ using Eqns (\ref{g}), (\ref{qr}) and (\ref{vm}) so
\begin{eqnarray}\label{evolution}
V_{m} &=& \tilde{G}_{m} R^{-1}_{m} C_{m}= J_{m-1} G_{m-1} R^{-1}_{m} C_{m}\nonumber \\
&=&J_{m-1} V_{m-1} C^{-1}_{m} R^{-1}_{m} C_{m}
\end{eqnarray}  
The key step is that if we choose $C^{-1}_{m} R^{-1}_{m} C_{m} = I$, then time evolution of $V$ becomes covariant as $V_{m}=J_{m-1} V_{m-1}$ and  $C^{-1}_{m} R^{-1}_{m} C_{m} = I$ implies that the time evolution of the $C$ matrix is generated backwards in time by the $R^{-1}$ matrices using
\begin{equation}\label{coeff}
C_{m-1} = R^{-1}_{m} C_{m}
\end{equation}   
   It is usual to normalize the columns of the coefficient matrix $C$ at each step so that the covariant vectors are unit vectors.
   

\section{The Quasi-one-dimensional System}

\begin{figure}[htbp]
	\begin{center}
		\includegraphics[scale=0.5]{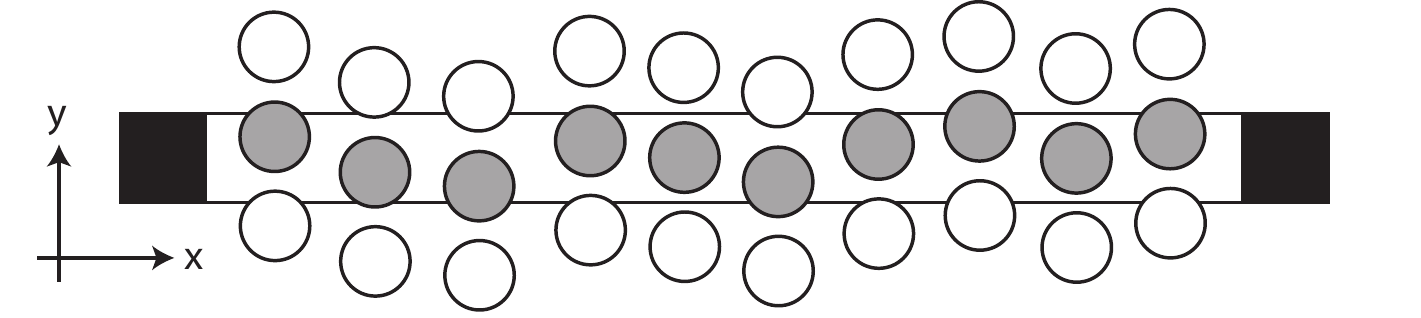}
			\caption{A schematic diagram of the Quasi One Dimensional $(H,P)$ system with hard wall boundaries in the $x$-direction and periodic boundaries in the $y$-direction. The shaded disks are within the QOD system and the unshaded disks are the first periodic images above and below each particle.}
		\label{QODsystem}
	\end{center}
\end{figure}

   The model we consider consists of $N$ hard disks (typically $80$) in a two-dimensional rectangular space $L_{x} \times L_{y}$.
   When $L_{y} < 2 \sigma$, where  $\sigma$ is the diameter of the disks, the space becomes a Quasi-One-Dimensional (QOD) system  \cite{TMg} and the hard disks remain ordered in the $x$-direction. 
   For the system considered here we use $L_y = 1.15 \sigma$, and scale $L_x$ according to the density $\rho = N \sigma^{2} /(L_x L_y)$ with hard-wall boundary conditions in the $x$-direction and periodic boundary conditions in the $y$-direction ((H,P) boundary conditions) see Fig. \ref{QODsystem}. 
   The phase space vector $\Gamma = (q,p)$ consists of all particle positions $q=({\bf q}_{1},...,{\bf q}_{N})$ and momenta $p=({\bf p}_{1},...,{\bf p}_{N})$ so the time evolution is composed of two components; the free-flight of particles between collisions and the change in momentum at collision \cite{CTM10}.
   The significant advantage of using the QOD system is that both the Lyapunov exponents and the Lyapunov modes of the system can be obtained to high accuracy by standard numerical schemes with fast convergence rates. 
   The exponents and mode structure for this system is well known consisting of a zero sub-space (where $\lambda=0$), and a number of transverse (T) and longitudinal-momentum proportional (LP) modes associated with the smallest positive and negative exponents \cite{MT09}. 

   The particle dynamics is symplectic \cite{AM78} so the Gram-Schmidt (GS) Lyapunov vectors $\delta \Gamma = (\delta q, \delta p )^{T}$ exhibit a characteristic structure.
   If the vector corresponding to the positive exponent is given by
\begin{equation}\label{conjugate1}
\delta \Gamma_{j} = \left( \begin{array}{c} \delta q \\ \delta p \end{array} \right),
\end{equation}
then the Lyapunov vector for the corresponding conjugate negative exponent is given by
\begin{equation}\label{conjugate2}
\delta \Gamma_{4N+1-j} = \left( \begin{array}{c} -\delta p \\ \delta q \end{array} \right).
\end{equation}
   These conjugate relations are obeyed by the GS vectors and as they imply orthogonality, they do not apply to the covariant vectors which  are in general not orthogonal but follow the directions of the stable and unstable manifolds.
   In the mode space the GS vectors occur in conjugate pairs and the majority of the dynamics occurs in the sub-space spanned by the pair of conjugate vectors \cite{MT09, MT11a}. 
   This together with a detailed knowledge of the tangent space dynamics allows us to quantitatively understand the angle distributions between conjugate modes \cite{TM11b}.
 

\section{Lyapunov Exponents and Modes}

\begin{figure}[htbp]
\begin{center}
	\includegraphics[scale=0.5]{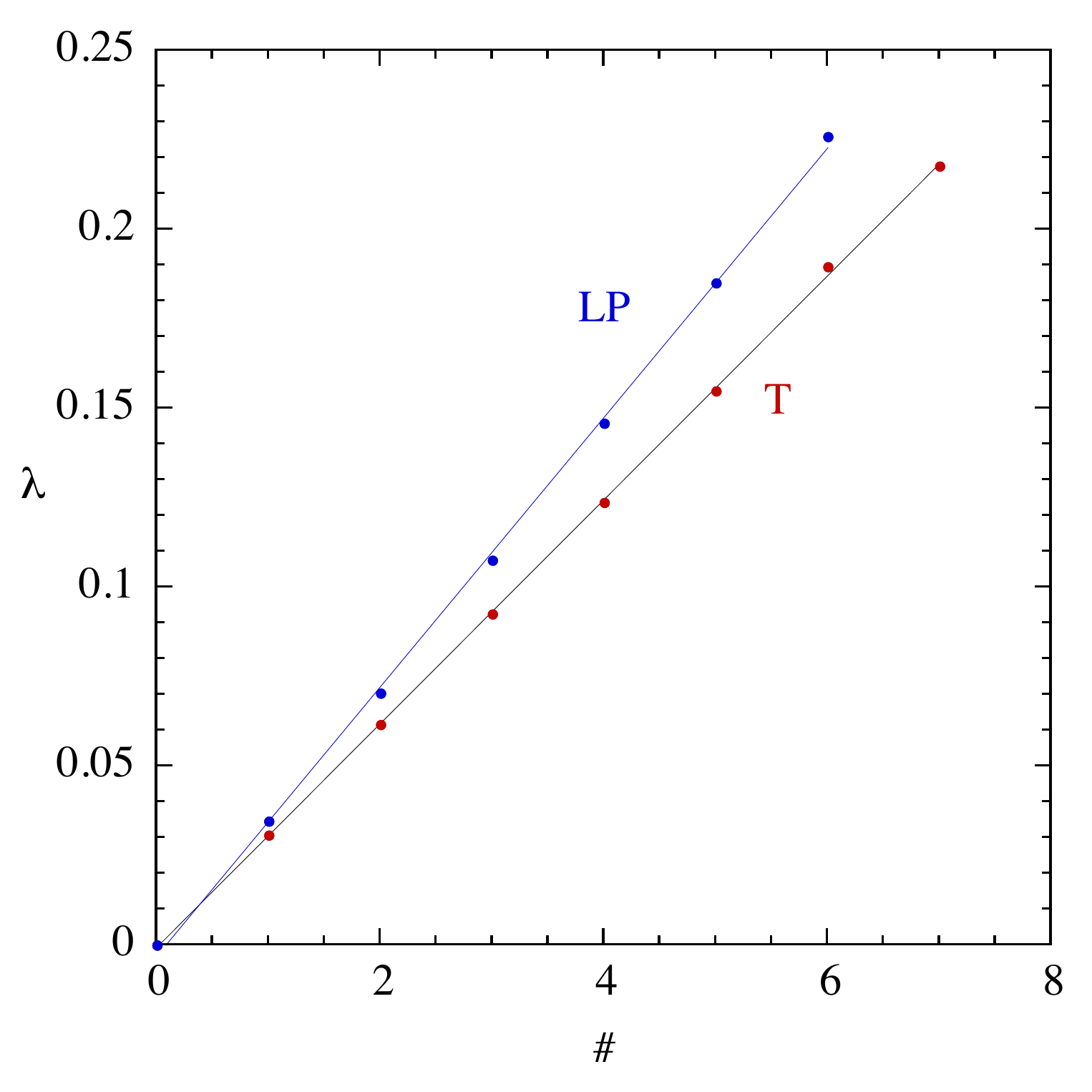}
\caption{(color online) The values of the Lyapunov exponents for the LP (blue) and T (red) modes of a system of $N=80$ hard disks at a temperature of $T=1$ and density $\rho=0.5$ plotted as a function of mode number. To a very good approximation the results are linear in mode number for both types of modes, with in this case a slope of $0.037654$ for T modes and a slope of $0.031295$ for LP modes.}
\label{T_LP_exp}
\end{center}
\end{figure}

\begin{figure}[htbp]
\begin{center}
	\includegraphics[scale=0.5]{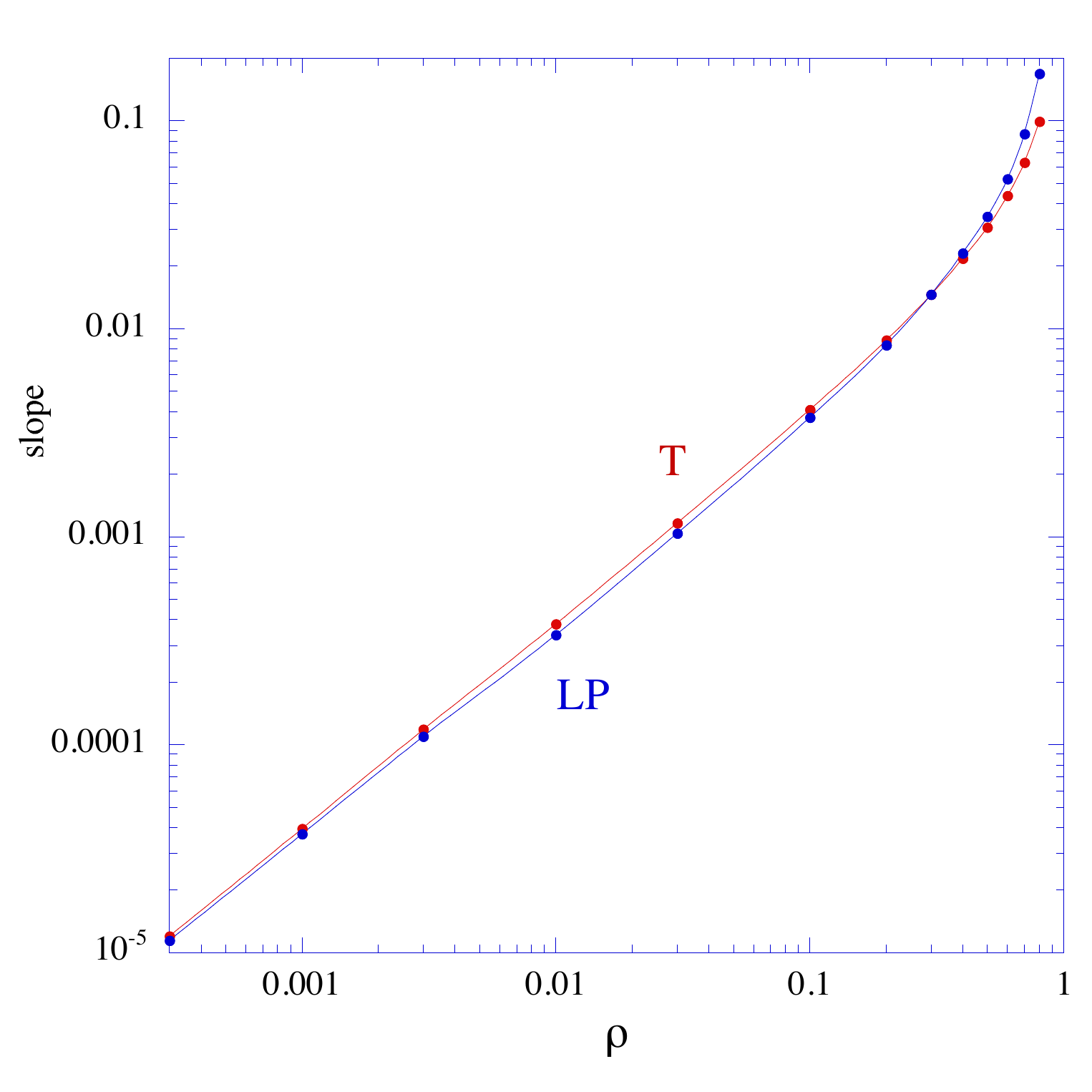}
\caption{(color online) The slopes of the linear fits in Fig. \ref{T_LP_exp} for systems of $N=80$ hard disks at a temperature of $T=1$, as a function of density $\rho$. For the low density or kinetic region the slope for the LP (blue) and T (red) modes are almost the same over a large density range suggesting stronger scaling properties. However, at $\rho=0.3$ the slopes cross over so that the LP mode has the largest slope.}
\label{T_LP_slopes}
\end{center}
\end{figure}

   It is well known \cite{TMg,Eck05} that the Lyapunov spectrum for the QOD system has exponents that appear in steps and associated with each of these exponents is a Lyapunov vector (or mode) that is delocalised and has one of two possible structures, either a transverse (T) mode or a longitudinal-momentum proportional (LP) mode.
   Occasionally, when the exponents corresponding to a T mode and a LP are equal (or approximately equal) the modes mix and become a time dependent linear combination of both mode forms (we refer to this as mode mixing) \cite{MT09}.
   However, two features of the spectrum remained unexplained. First the size of the step region in the exponent spectrum, and second the order in which the modes appear for a particular density.
   Recent calculations for systems of $100$, $150$ and $200$ particles suggest that the thermodynamic limit for the spectrum has been reached at $N=100$ and that a fixed fraction of the spectrum consists of steps (and thus modes) \cite{MT11a}.
   The order in which modes appear can also be determined from Fig. \ref{T_LP_slopes}.
   For any fixed density, the exponents when plotted as a function of T mode number or LP mode number are to a good approximation linear with slopes that depend upon the density, as shown in Fig. \ref{T_LP_exp} for a density of $\rho=0.5$.
   Once the slopes of these lines are determined for a particular density by reading them off Fig. \ref{T_LP_slopes}, graphs like that in Fig. \ref{T_LP_exp} can be constructed and the T and LP exponents read off.
   The order of the steps is then determined by simply ordering the exponents from smallest to largest.

   The slopes of the linear fits for the T and LP modes are shown in Fig. \ref{T_LP_slopes}  as a function of density for $0.0003 <\rho < 0.8$.
   We see immediately that for densities below $0.3$ the order of the steps is fixed, first an LP mode then a T mode, and then that pattern repeats throughout the step region. 
   In fact this region of densities $\rho < 0.1$ has already been termed the {\it kinetic region} due to the behaviour of the largest exponent and the localization  \cite{TM03b}. 

   The functional form of GS and covariant modes are largely the same for the modes corresponding to positive exponents $n>0$ but the modes corresponding to negative exponents are different.
   The negative covariant vectors are not orthogonal to their conjugate vectors and therefore there is a well defined linear combination of modes which they comprise. 
   
\subsection{The Lyapunov Exponent Spectrum}

\begin{figure}[htbp]
\begin{center}
	\includegraphics[scale=0.5]{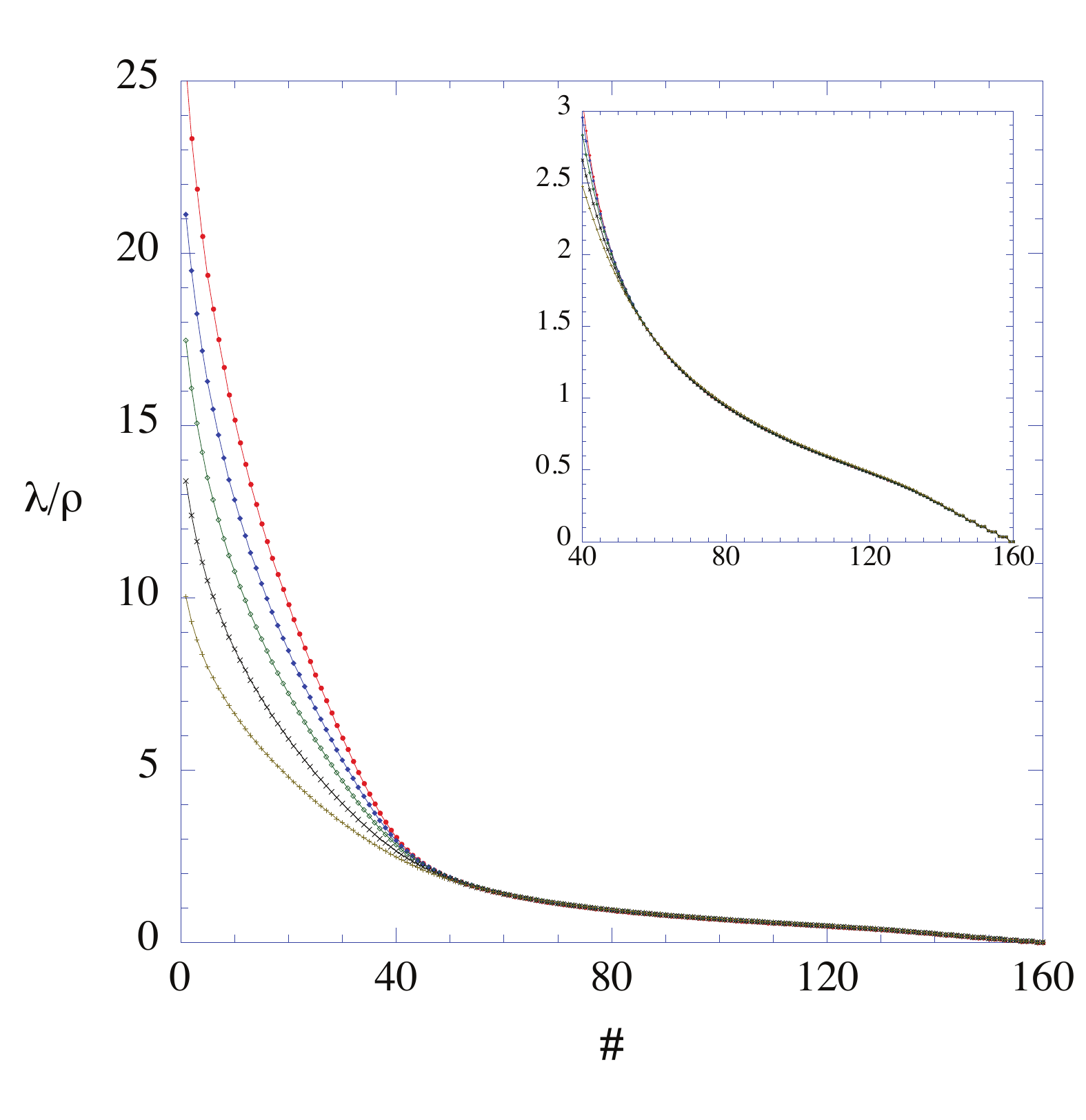}
\caption{(color online) The scaled Lyapunov spectrum ($\lambda / \rho$) for the positive exponents in the {\it kinetic regime} for systems of $N=80$ hard disks at a temperature of $T=1$ (that is,  densities ranging from $\rho=0.0003$ to $\rho=0.03$). The modes occur in the same order throughout this regime, with an LP mode followed by a T mode. Notice that for all densities below $\rho=0.1$ the values of $\lambda_{n} /\rho$ are the same for exponent numbers $40 \le n \le 160$. In contrast, the scaling is not linear in density for the first $40$ exponents.}
\label{rho_scaling}
\end{center}
\end{figure}

   A systematic approach to the study of the Lyapunov spectrum as a function of density begins by looking for scaling behaviour in the exponents.
   For the QOD system it is known that the largest exponent at densities below $\rho \sim 0.1$ depends on the density as $\alpha \rho \ln (\beta \rho)$ as shown by Krylov \cite{Kry79} and this has been confirmed in previous simulation studies for the QOD system  \cite{TM03b} and more generally for two-dimensional systems of disks \cite{Del96}.
   As the Lyapunov spectrum for this system has been studied before we limit our description to features not previously observed. 
   In Fig. (\ref{rho_scaling}), by plotting $\lambda/\rho$ against exponent number, we observe that for a large proportion of the Lyapunov spectrum in  the {\it kinetic region}, that is, exponent numbers between $40$ to $160$, Lyapunov exponents scale with density $\lambda \propto \rho$ for the $N=80$ disk system.
   This is observed in the whole positive half of the spectrum, except for the first $40$ exponents (for $N=80$), and includes all of the step region where the corresponding Lyapunov vectors are modes.
   It is  known that the largest exponent scales as $-\rho \ln \rho$ in the kinetic region so it is reasonable to assume that all exponents that are not in the linear region (that is $2$ to $40$) will scale in the same way as the largest.
   The proportion of the spectrum composed of steps and Lyapunov modes appears to remain fixed as the system size increases \cite{MT11a} so we expect that the scaling of the KS entropy is dominated by the scaling behaviour of the first $40$ exponents.

\section{Scalar Properties}

   The thermodynamic and dynamical behaviour of hard disks in narrow channels has been studied before \cite{FMP04} as a function of $L_{y}$.
   This system becomes QOD  when $L_{y}<2$ and it is found that the collision frequency and the Lyapunov spectrum scale strongly with the width of the system $L_{y}$.
   This implies a strong scaling of both the largest exponent and the KS entropy with $L_{y}$.
   

   If a system of $N$ particles has $N_{t}$ collisions in time $t$ then the single particle collision frequency is given by $v_{2}=2n_{t}/(Nt)$.
   For the QOD system the single particle collision frequency is directly related to the potential contribution to the pressure as
\begin{equation}
v_{2} = \frac {2(Pv - kT)} {\Delta p}
\end{equation}
where $\Delta p$ is the average momentum transfer per collision.
 Here we see that the single particle collision frequency at a density of $0.01$ changes by a factor of $3$ as $L_{y}$ goes from $1$ to $2$, that is over the range of $L_{y}$ values that correspond to the QOD system. 
   

   There have been a number of theoretical studies of the KS entropy per particle of hard-particle systems in two and three dimensions \cite{vB97}.
   In the low density limit the KS entropy is expected to be linear in the  single-particle collision frequency $\nu_{2}$ given by
\begin{equation}
\frac {h_{KS}} {N} = \nu_{2} A ( - \ln \rho + B +O(\rho) )
\end{equation}
where $A$ and $B$ are constants, and will thus change dramatically with the width of the QOD system $L_{y}$. 
   As the KS entropy is simply the sum of the positive Lyapunov exponents its scaling behaviour is determined by the dominant scaling of individual groups of exponents, in particular the first $40$ exponents for a $80$ disk system,
   As we have only considered one fixed value of $L_{y}=1.15$, we have not attempted a discussion of the strong scaling of these properties with $L_{y}$ .


\section{Localization}

\subsection{Order Parameter}
 
   The contribution of particle $i$ to the $j^{th}$ Lyapunov vector $\delta \Gamma^{(j)}$ can be written as $\chi^{(j)}_{i} =\left(  \delta q^{(j)}_{i} \right)^{2} + \left( \delta p^{(j)}_{i} \right)^{2}$. 
   By definition $\chi^{(j)}_{i}$ is positive and as the Lyapunov vector is normalised, $\chi^{(j)}_{i}\le 1$ for each particle.
   Therefore, for each Lyapunov vector $j$ the set $\{\chi^{(j)}_{i}\}$ are positive random variables which sum to one. 
   It is natural to construct an entropy function for each Lyapunov vector \cite{TM03b} from the set $\{\chi^{(j)}_{i}\} $ and this gives a measure of the localization of the Lyapunov vector 
\begin{equation}\label{localization}
W_{N}^{(j)} (t) =\frac {1} {N} \exp \left( - \sum_{i=1}^{N} \chi^{(j)}_{i} \ln \chi^{(j)}_{i} \right).
\end{equation}
   This measure $W^{(j)} (t)$ is close to zero for a highly localized vector and close to one for a strongly delocalized vector. 
   It gives an instantaneous measure of the localization for each vector $j$ and represents the proportion of the $N$ particles that contribute to the Lyapunov vector. 
   Indeed $N W_{N}^{(j)}$ could be considered the {\it participation number} - the number of contributing particles - which is independent of $N$ if the vector is strongly localized on a subset of the particles.
   In this way it relates more directly to the {\it inverse participation ratio} $Y_{2}$ \cite{Man85, Kan86, Cha93} which is also used as a measure of the number of effective degrees of freedom.
   However, no simple exact relation between $N W_{N}^{(j)}$ and $Y_{2}$ can be given.

   In what follows, we will investigate all aspects of the localization for the QOD system - the average localization of both GS and covariant vectors, the probability distribution for the localization, configurations which lead to strong localization, instantaneous localization and the connections between localization and angle distributions.
      
\subsection{Average Localization}

\begin{figure}[htbp]
\begin{center}
	\includegraphics[scale=1.0]{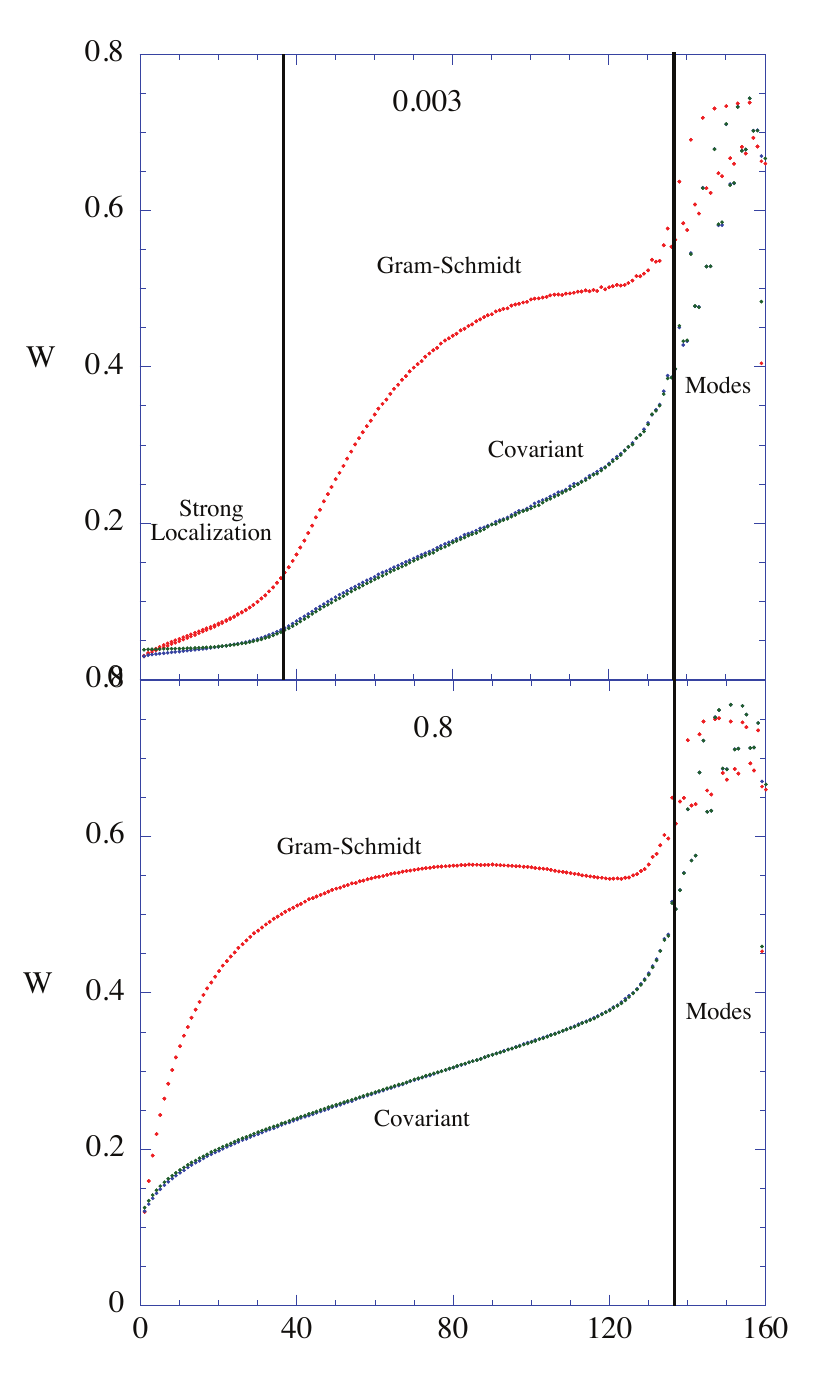}
\caption{(color online) The average localizations $W$ for both the GS (red) and the covariant (blue) vectors of a QOD system of $80$ hard disks at a temperature of $T=1$, at two different densities. In the first figure the density is $\rho=0.003$ and for the second figure the density is $\rho=0.8$. There is clear evidence of strong localization in the low density state for vector numbers $j < 35$. In the mode region the localization splits into two branches, one for the T modes and the other for the LP modes.}
\label{average_loc}
\end{center}
\end{figure}

   The average localization for a QOD system of $N=80$ disks at two different densities is given in Fig (\ref{average_loc}).
   One density $\rho=0.003$ is typical of the kinetic region and the other $\rho=0.8$ is typical of the dense liquid phase. 
   It has been observed \cite{TM03b} that the localization of the GS vectors as a function of density generally takes one of two forms: one typical of a low density kinetic region, and the other typical of a  dense liquid.
   Strong localization occurs in the kinetic region when the density is less than $0.1$, and here a group of Lyapunov vectors associated with the largest exponents show small values of $W_{N}^{(j)}$. 
   Indeed, at very low densities a number of vectors all show a similar form of strong localization which appears to depend linearly on the vector number \cite{TM06} in the localization spectrum.
    For the GS vectors the conjugate relation Eqn (\ref{conjugate2}) implies that the localizations for the two conjugate vectors are identical and this is indeed what is observed in Fig. (\ref{average_loc}).
   The behaviour of the localization for the range of vectors $140<j<160$ (depending on system size) is associated with the different localization values for T modes and LP modes. 
    The GS localization spectrum at high density ($\rho=0.8$) is essentially exponential in the vector number and more delocalized than the covariant vectors which look linear in the vector number.
    At low density ($\rho=0.003$), as a function of vector number, the localization is initially linear before returning to the usual exponential shape.
   In Fig (\ref{average_loc}) we see the same effect at a density of $\rho = 0.003$ where both the GS and covariant localizations depend linearly on the vector number between $1$ and about $35$.
   The randomly dropped brick model of \cite{TM06} gives an estimate of the number of most localized GS vectors that can co-exist and not violate the orthogonality condition.
   For $80$ disks it is possible to have $40$ such vectors but on average random positioning reduces this to approximately $35$.
   This is in good agreement with the results reported in Fig. (\ref{average_loc}).
   
   At high density the covariant modes with positive exponents have similar localizations to those associated with negative exponents, but at low density there are small differences.
   The localization of the Lyapunov vectors associated with the most negative exponents appears to be different to the localizations of the conjugate vectors, suggesting the possibility of some convergence issues.
   The Gram-Schmidt vectors, from which the covariant vectors are calculated, satisfy an exact conjugate relation for the localization however, the linear combinations which make up the covariant vectors need not.
   However, the time reversal symmetry of this system suggests that the localizations for positive and negative vectors should be the same.

   The covariant vectors at low density have a localization spectrum that is again linear but with smaller slope (almost constant).
   For covariant Lyapunov vectors there is no longer a strict orthogonality constraint so subsequent Lyapunov vectors can contain a significant component of the first, second or higher, Lyapunov vectors.
   Therefore the localization characteristics of the first covariant vector can influence the localization of subsequent covariant Lyapunov vectors more strongly.
   The $j^{th}$ covariant vector can be written as a linear combination of the first $j$ GS vectors as $v_{j} = \sum_{i=1}^{j} c_{i} g_{i}$ where $c_{i}$ are coefficients, so for $v_{j}$ to be localized we require all of $g_{1},...,g_{j}$ to be localized on essentially the same group of particles.
   Even if all component vectors are strongly localized it does not necessarily follow the the covariant vector is localized, although the numerical evidence suggests that the same number of strongly localized vectors occurs in both the GS basis and the covariant vectors, the covariant ones are on average more localized than the GS vectors.

\subsection{Localization Distributions}
     
\begin{figure}[htbp]
\begin{center}
	\includegraphics[scale=0.8]{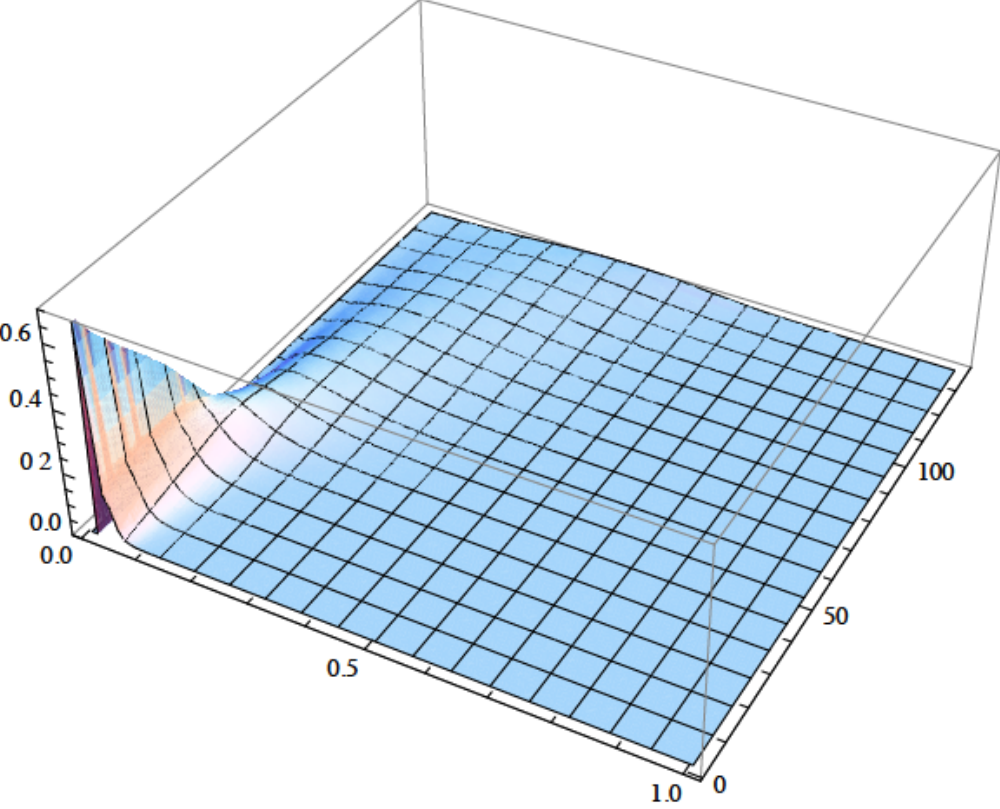}
\caption{(color online) The probability distribution of the localization $W$ for the covariant vectors of a QOD system of 80 hard disks in the {\it kinetic} regime at a density of $\rho=0.003$ and temperature of $T=1$. The axis labeled $0.0$ to $1.0$ is the value of the localization, the axis labeled $0$ to $127$ is the vector number and the vertical axis is the probability. The most probable localization value for the first $35$ vectors is the minimum localization of $2/N$. There is also a small cusp at $4/N$ corresponding to a linear combination of two most strongly localized configurations. For vectors larger than $40$ the localization distributions are smooth and very similar to those of higher density systems where the distribution has a smooth maximum.}
\label{loc_prob_003}
\end{center}
\end{figure}

\begin{figure}[htbp]
\begin{center}
	\includegraphics[scale=0.8]{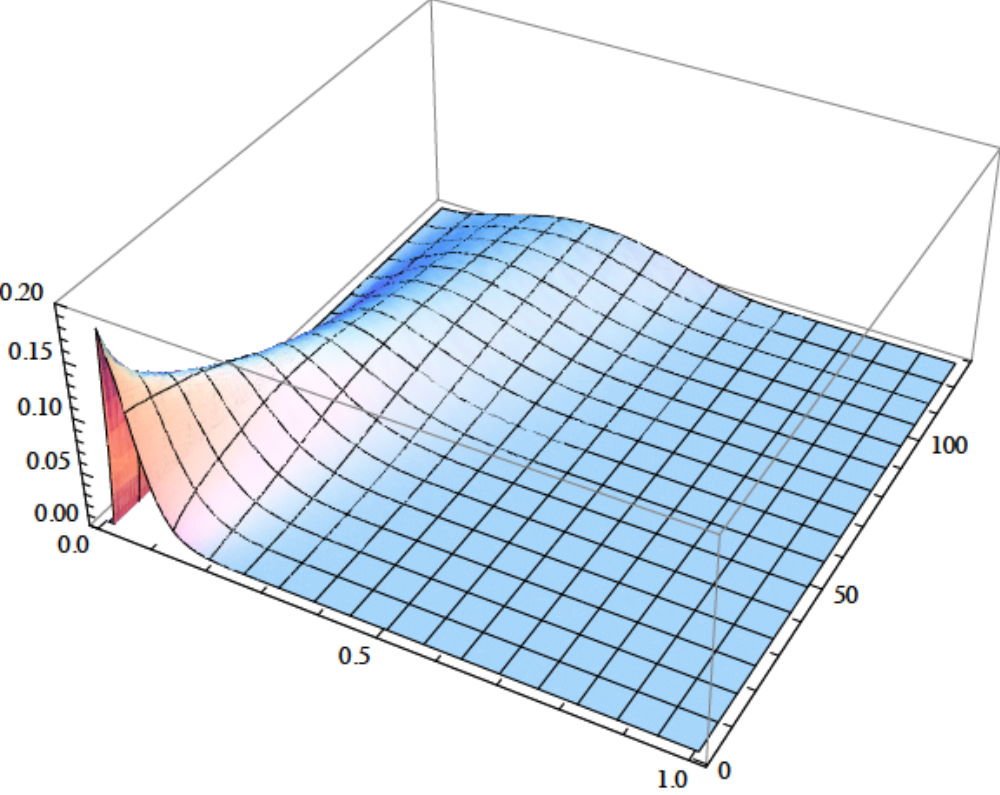}
\caption{(color online) The probability distribution of the localization $W$ for the covariant vectors of a QOD system of 80 hard disks at a density of $\rho=0.8$ and temperature of $T=1$. The figure is to be compared with the kinetic region in Fig (\ref{loc_prob_003}) but here the vertical scale is one third smaller. Otherwise the axis labels are the same as Fig (\ref {loc_prob_003}). There is a small remnant of the most localized configuration in the first few vectors.}
\label{loc_prob_8}
\end{center}
\end{figure}

   The localization $W_{N}^{(j)} (t)$ is defined at each instant in time so we can construct a probability distribution for this quantity for each Lyapunov vector in Figs. (\ref{loc_prob_003}) and (\ref{loc_prob_8}).
   This gives more detailed information than the average localization $<W_{N}^{(j)}>$ in Fig. (\ref{average_loc}).
   The average localization as a function of density shows that the zero modes and the T and LP modes are only weakly density dependent. 
   The major feature of the localization in the continuous part of the spectrum is that the vector corresponding to the largest exponent becomes more and more localized with decreasing density or increasing $N$.

   In Fig. (\ref{loc_prob_003}) the distribution of localization for the first $127$ Lyapunov vectors at a low density ($\rho = 0.003$) for a QOD system of $80$ disks are shown.
   Initially there is a dominant contribution at the position of the most localized configuration $W=2/N$ with a barely visible blip at $W=4/N$ corresponding to four equal contributions to the set $\{ \chi_{i} \}$.
   However, with increasing vector number the most localized peak diminishes apparently linearly and the distribution becomes broad and smooth.
   For Gram-Schmidt vectors this most localized contribution occurs only for the first quarter of the expanding Lyapunov vectors, but for the covariant Lyapunov vectors the influence of this most localized contribution can contribute to more vectors as seen in Fig. (\ref{loc_prob_003}).
   Therefore the structure of the most localized configuration plays an even more important role for the covariant Lyapunov vectors.
   At high density there is a remnant of the strong localization configuration evident in Fig. (\ref{loc_prob_8}) but its amplitude is only $1/3$ of that at low density and decays more quickly than linearly with vector number, and with the scale change in the graph the distribution with a smooth maximum is more apparent.

\subsection{Strongly Localized Configurations}

   In Fig. (\ref{loc_prob_003}), for each of the first $35$ Lyapunov vectors, there is clear evidence of an instantaneous localization of $W=2/N=0.025$ which corresponds to a configuration with $\chi =0.5$ on two neighbouring particles remaining for long periods of time.
   The position of the localized pair of particles is not important but in the transition of the localization from one pair of particles to another a variety of things can happen.
   The first possibility is that the localization on two particles decays from an initial value of $1/2$ to zero as the localization on a different two particles increases from zero to $1/2$.
   This behaviour is clear in the first GS vector (in Fig. (\ref{loc_prob_003})) between collision numbers $80$ and $90$ where there is a plateau value of $4/N$ as predicted above.
   There are other possibilities as we see localization values above $4/N$ for short periods which involve more than 4 particles, and other smaller blips as the pair of vectors moves one place in either direction.
   It is evident that at this time the second covariant vector is essentially equal to the first GS vector (although it must contain some small component of the second GS vector).

\begin{figure}[htbp]
\begin{center}
	\includegraphics[scale=0.3]{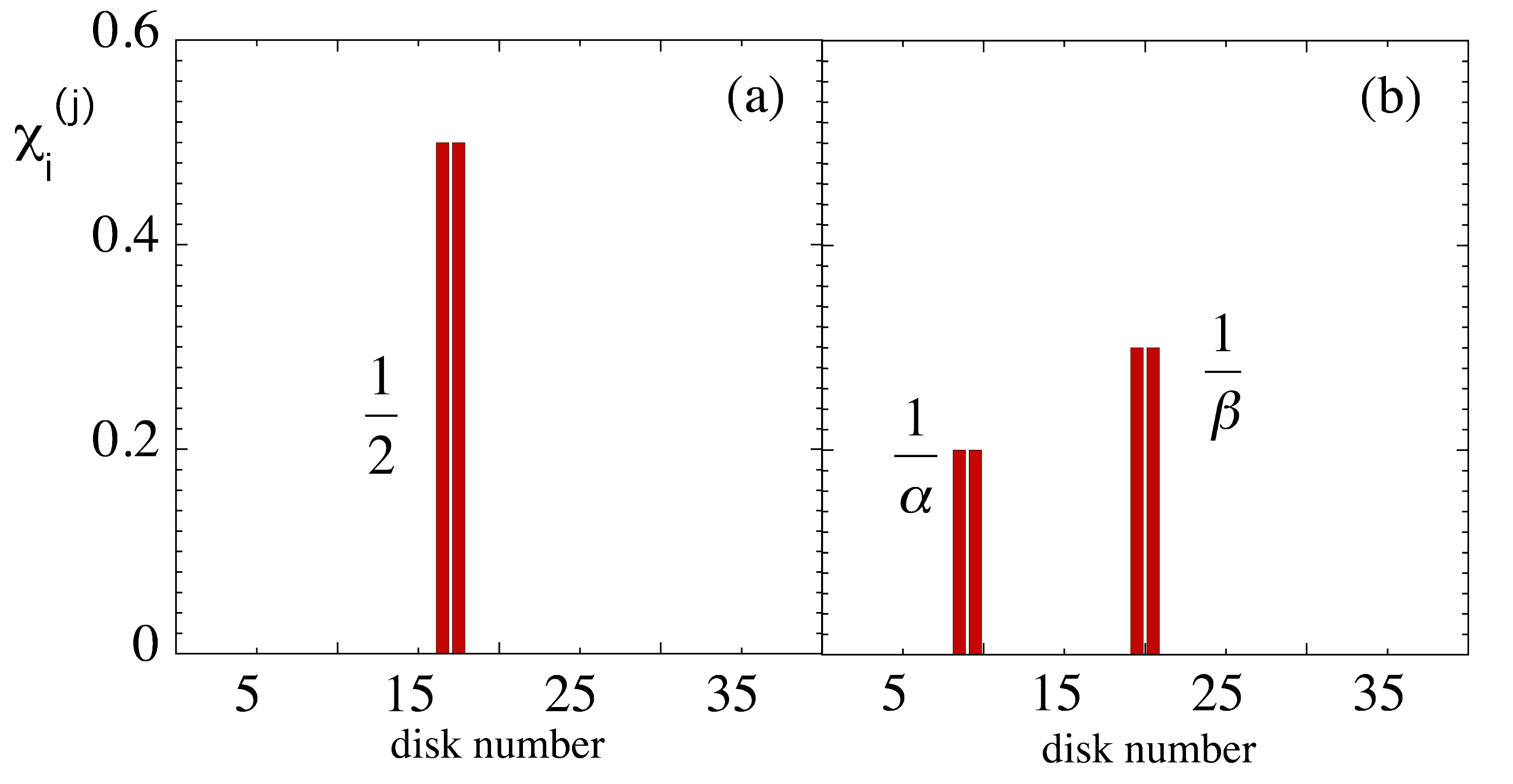}
\caption{(color online) (a) The configuration which gives the strongest localization $W^{(j)}$ for the QOD system of hard disks. Here $\chi^{(j)}_{i} =0.5$ for $i=17$ and $i=18$ and thus this configuration gives $W^{(j)}=2/N$. (b) A linear combination of two of the most localized configurations gives a range of values of localization depending upon the particular linear combination. If $\chi^{(j)}_{i} = 1/\alpha$ for $i$ and $i+1$ and $\chi^{(j)}_{k} = 1/\beta$ for $k$ and $k+1$ the localization depends on the values of $\alpha$ and $\beta$ which are not independent. When $\alpha=0$, $W_{N}^{(j)}=2/N$  and when $\alpha = \beta$ then $W^{(j)}=4/N$.}
\label{loc_conf}
\end{center}
\end{figure}

    The most localized configuration is when $\chi^{(j)}_{i} = 0.5$ for both $i$ and $i+1$ and is zero for all the other values of $i$ (see Fig. \ref{loc_conf} (a)), then the localization measure is given by $W^{(j)} (t)= 2/N$.
   Clearly the vector is localized on only two particles so that $W^{(j)}$ follows a strict power law in the number of particles $N$, with the power equal to $-1$.
   Another possible strongly localized configuration is a mixture of two of these most localized vectors (see Fig \ref{loc_conf} (b)).
   As this is a mixture, we expect that  $\chi^{(j)}_{i} = 1/\alpha$ for $i$ and $i+1$ 
and for $\chi^{(j)}_{k} = 1/\beta$ for $k$ and $k+1$ where $\beta = \alpha / (\alpha -1)$.
   The localization for this mixture is $W^{(j)} (t)= \alpha (\alpha -1)^{(1/\alpha -1)}/N$ which has a maximum when $\alpha = 4$.
   In fact this combination only gives localization values between $2/N$ and $4/N$.
   

\subsection{Instantaneous Localization}

\begin{figure}[htbp]
\begin{center}
	\includegraphics[scale=0.25]{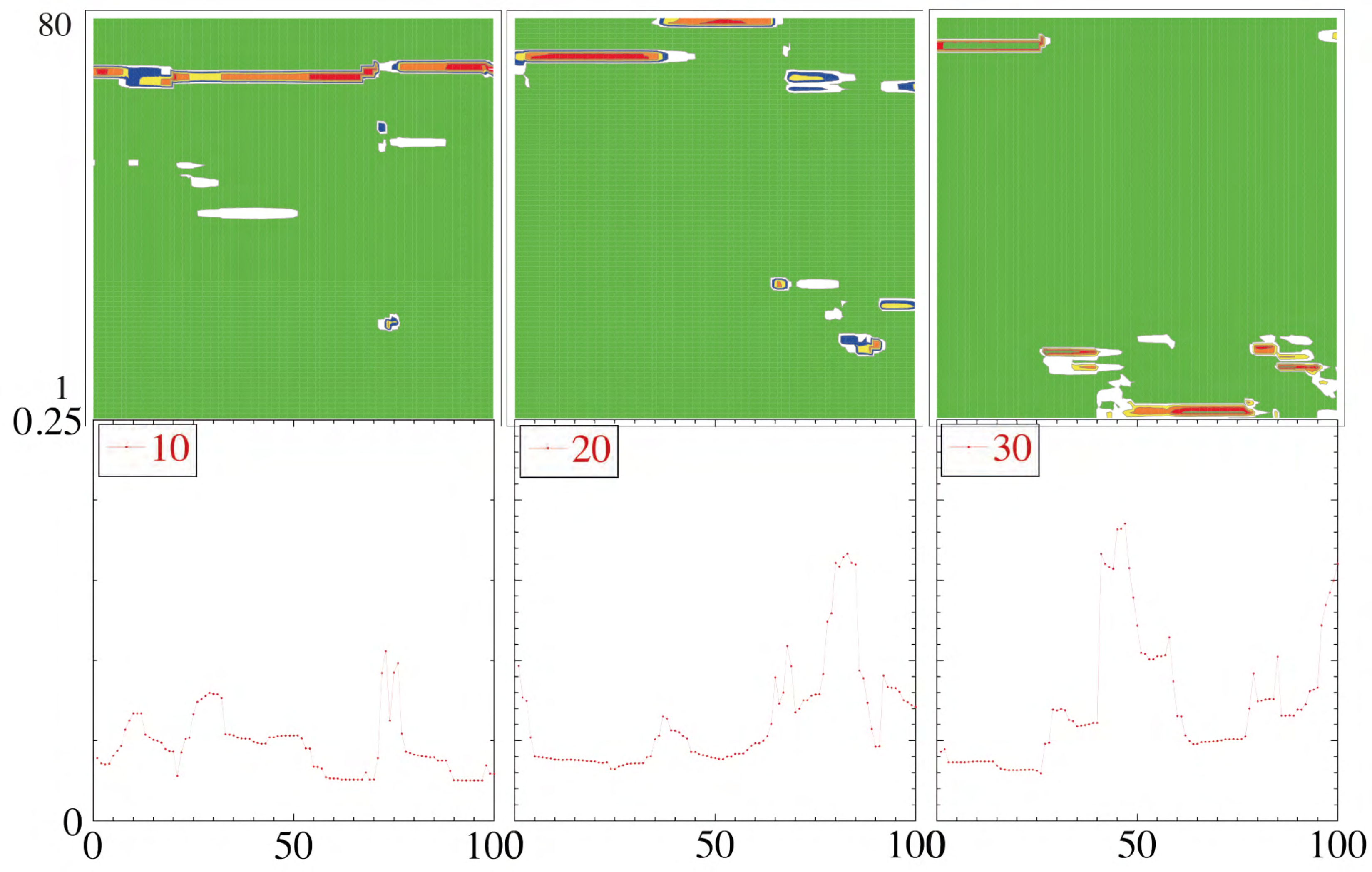}
\caption{(color online) The localization contributions from each particle $\chi_{i}^{(j)}$ calculated as a function of collision number for a selection of the GS vectors of a QOD system of $N=80$ hard disks at low density $\rho=0.003$ and a temperature of $T=1$. The upper panels are the values of $\chi_{i}^{(j)}$ for each particle for Lyapunov vectors $j=10, 20, 30$ with the lower panels giving the instantaneous value of the localization for the system $W(t)$. The vertical axis on the upper panels is the particle number and on the lower panels is the localization. The horizontal axis is the collision number for all graphs. The color changes on the upper panels correspond to levels $0.05, 0.15, 0.25, 0.35$ and $0.45$ and the corresponding colors are in order: green, white, blue yellow, orange and red. The vector $j=10$ is strongly localized and for several periods its value is $2/N=0.025$, so its configuration corresponds to Figure (\ref{loc_conf}). The localization is less strong for vectors $j=20$ and $j=30$. }
\label{GS_loc_102030}
\end{center}
\end{figure}

   In Fig. (\ref{average_loc}) for $N=80$, the low density average localization for the covariant Lyapunov vectors shows an initial strongly localized region associated with the first $35$ Lyapunov vectors.
   A similar behaviour has been seen previously in the average localization of the GS vectors and there it was argued that each of these vectors was dominated by the most localized configuration in Fig. (\ref{loc_conf}) and the required orthogonality limited the number of vectors.
   Here we know that the covariant vectors are linear combinations of the GS vectors, but that in general the covariant vectors are more localized than the equivalent GS vectors.
   Here we look at the dynamics of the localization in order to understand this property of the covariant vectors.
   In Fig. (\ref{GS_loc_102030}) we show the evolution of the $\chi_{i}^{(j)}$ for each particle, from the GS vectors, over $100$ collisions and at the same time the evolution of the localization order parameter $W(t)$.
   For vector $10$ there are two places where we see the most localized configuration around $60$ and $90$ collisions but there are larger excursions that do not correspond to either of the two strongly localized configurations.
   Similarly, for vectors $20$ and $30$ there seem to be no most localized configurations and increasingly larger excursions in $W_{N}(t)$.
   
\begin{figure}[htbp]
\begin{center}
	\includegraphics[scale=0.25]{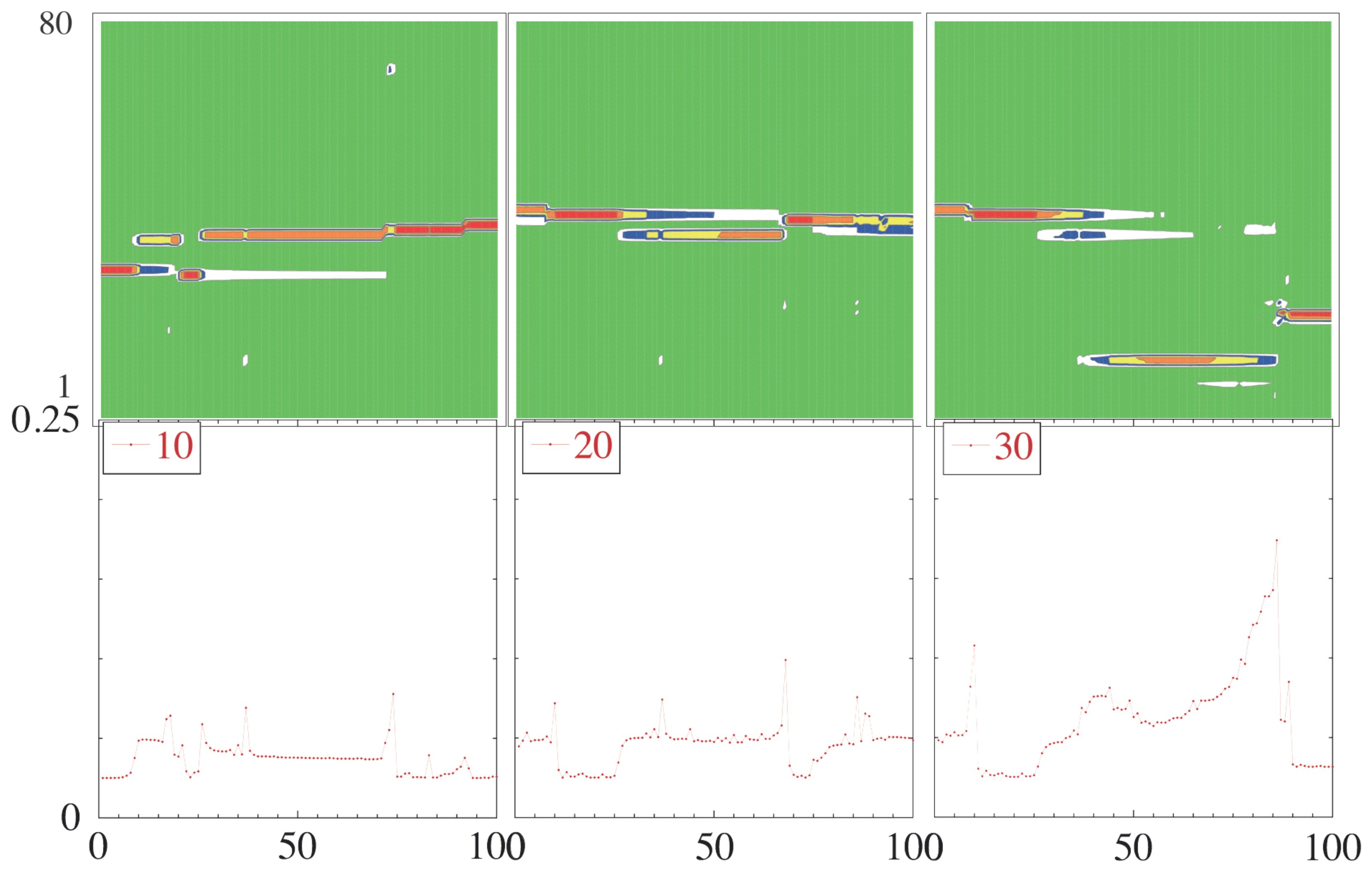}
\caption{(color online) The localization contributions from each particle $\chi_{i}^{(j)}$ calculated as a function of collision number for a selection of the covariant Lyapunov vectors of a QOD system of $N=80$ hard disks at low density $\rho=0.003$ and a temperature of $T=1$. The upper panels are the values of $\chi_{i}^{(j)}$ for each particle for covariant Lyapunov vectors $j=10, 20, 30$ with the lower panels giving the instantaneous value of the localization for the system $W(t)$.  The axes and color changes are same as in Fig. (\ref{GS_loc_102030}). There is evidence of strong localization in both $j=10$ and $j=20$, but for $j=30$ the localization is less strong. }
\label{C_loc_102030}
\end{center}
\end{figure}

  In Fig. (\ref{C_loc_102030}) we show the same time evolution of $\chi_{i}^{(j)}$ for each particle, for the covariant vectors, over the same $100$ collisions.
  Vector $10$ is initially localized on two particles for the first $\sim10$ collisions so $W(t)=2/N=0.025$ and then quickly splits into two two particle pairs where $W(t)\sim0.05$.
  This remains until $\sim 75$ collisions when the localization returns to the value $W(t)=0.025$.
  The other two vectors $20$ and $30$ also show short regions where the strong localization $W(t)=0.025$ occurs but for vector $20$, $W(t)\sim0.05$ is more common and then at vector $30$ larger excursions from the two most localized configurations involving more than $4$ particles are observed.
  
  The time evolution of $\chi_{i}^{(j)}$ for each particle, for the covariant vectors at a high density of $\rho=0.8$, over $100$ collisions shows that none of the vectors show the strongest localization but there are periods where the localization is principally on four particles.
  

\subsection{Strong Localization}

\begin{figure}[htbp]
\begin{center}
	\includegraphics[scale=0.5]{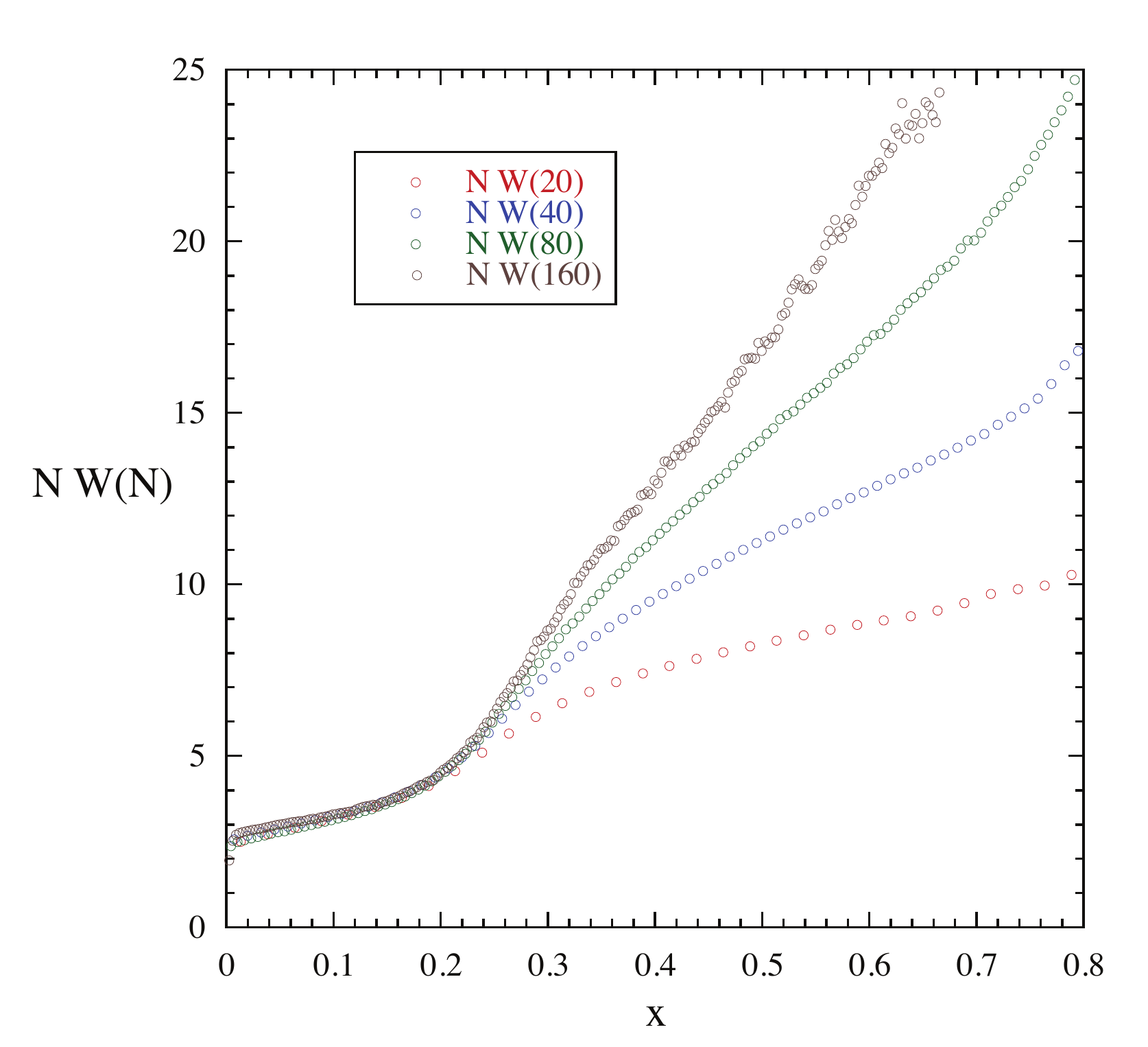}
\caption{(color online) The participation number $N \left <W_{N}^{(j)}\right >$ for the covariant vectors of a QOD system of hard disks as a function of the number of particles $N$ at low density ($\rho = 0.003$) and a temperature of $T=1$. The horizontal axis is a continuous variable constructed from the vector number $j$ using $x = (j-1/2)/2N$. There is evidence of strong localization in the first $20\%$ of the covariant vectors corresponding to positive exponents. As the system size is increased the same proportion of vectors remains strongly localized.}
\label{Astrong_loc_L}
\end{center}
\end{figure}

   To establish asymptotic strong localization in the thermodynamic limit we need to look at the localization as a function of $N$.
   If the system exhibits asymptotic strong localization then only a finite number of degrees of freedom are involved and this does not change with system size $N$.
   Localization on only two particles is the most localized configuration possible.
   Another common situation is when the localization is confined to a finite number of particles, strictly less then $N$.
   In that case the entropy in the exponent of Eqn (\ref{localization}) is a constant so the {\it participation number} $N W_{N}^{(j)} (t)$ which is constant independent of $N$. 
   In Fig. (\ref{Astrong_loc_L}) we plot the participation number for QOD systems of $N=20, 40, 80, 160$ particles, and the results show that the region of strong localization already identified in the average localization figures, is indeed strongly localized.

\subsection{Localization Scaling}

\begin{figure}[htbp]
\begin{center}
	\includegraphics[scale=0.5]{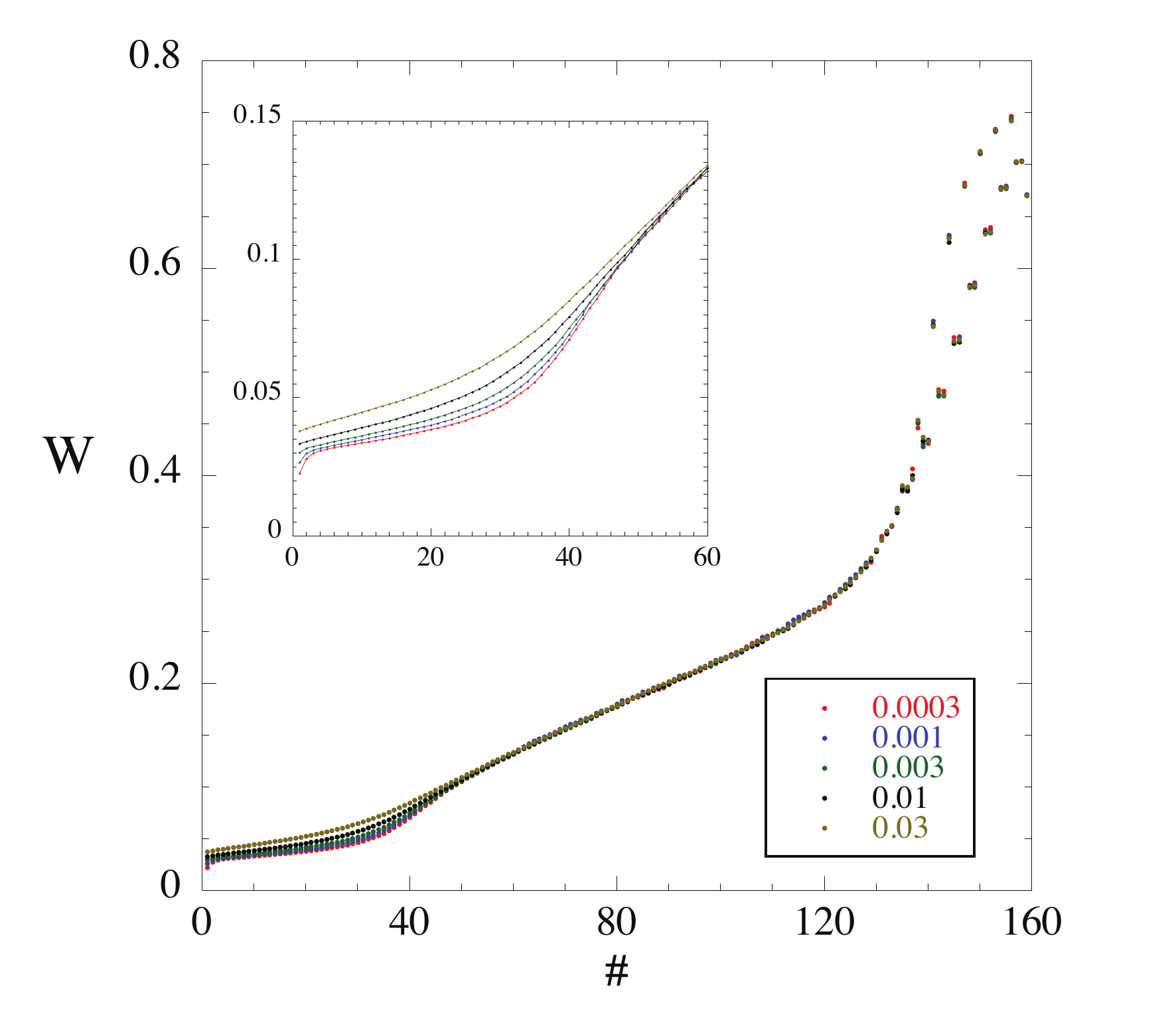}
\caption{(color online) The  localization spectrum ($W$) for the positive covariant Lyapunov vectors for systems of $80$ hard disks at a temperature of $T=1$ with densities in the kinetic regime ranging from $\rho=0.0003$ to $\rho=0.03$. The curve for $0.0003$ is red, $0.001$ is blue, $0.003$ is green, $0.01$ is black and $0.03$ is brown. Above $140$ the modes separate into two branches and alternate between the values for LP modes then T modes for all densities, and again the localization of each mode is independent of the density. For densities below $0.1$ the values of $W$ are the same for vectors in the range $45 \le n \le 160$. The first $50$ Lyapunov vectors show some small systematic variation in localization as evidenced in the inset. Here the localization becomes smaller with smaller density.}
\label{rho_scaling_W}
\end{center}
\end{figure}

   Earlier we saw that in the kinetic regime the Lyapunov spectrum exhibited a scaling relation with the exponents being linear in density. 
   Here we see that in the same range of exponent numbers, that is $45$ to $160$ for a $80$ disk QOD system, the average localization is a constant independent of density.
   The results are shown in Fig. (\ref{rho_scaling_W}), and the inset in that graph shows how the initial $45$ exponents do change with density.

\section{Angle Distributions}

   The distribution of angles between covariant Lyapunov vectors $j$ and $j+d$ is observed to be largely independent of $j$ but strongly dependent on $d$.
   A simple geometric argument has been proposed to explain this observation \cite{TC10}.
   Imagine placing an initial unit vector randomly directed at the origin of a high-dimensional space. 
   Placing a second vector at random can be achieved by choosing a circle of unit radius containing the head of the first vector, and choosing a random point on this circle as the end-point of the second vector.
    The angle distribution between these two vectors is uniform.
    For the third vector we consider a sphere containing the heads of vectors $1$ and $2$, and choose a random point on the sphere to be the head of the third vector.
    The resulting distribution is $\rho(\theta_{13})=\sin( \theta_{13})$.
    It follows that the distribution of angles between two vectors with indices differing by $d$ is $\rho(\theta_{d})=\sin^{(d-1)}( \theta_{d})$.
    Here we see that this geometric argument works reasonably well at high densities, but at low densities in the kinetic regime, the angle distributions are affected by the strong localization of some of the covariant Lyapunov vectors.
    Indeed, looking back at the high density results there still remains a small signature due to localization in the first few Lyapunov vectors.
   The angle $\theta_{ij}$ between a pair of Lyapunov vectors $i$ and $j$ is calculated at each collision and  a distribution function for the angle constructed.
    
\begin{figure}[htbp]
\begin{center}
	\includegraphics[scale=0.6] {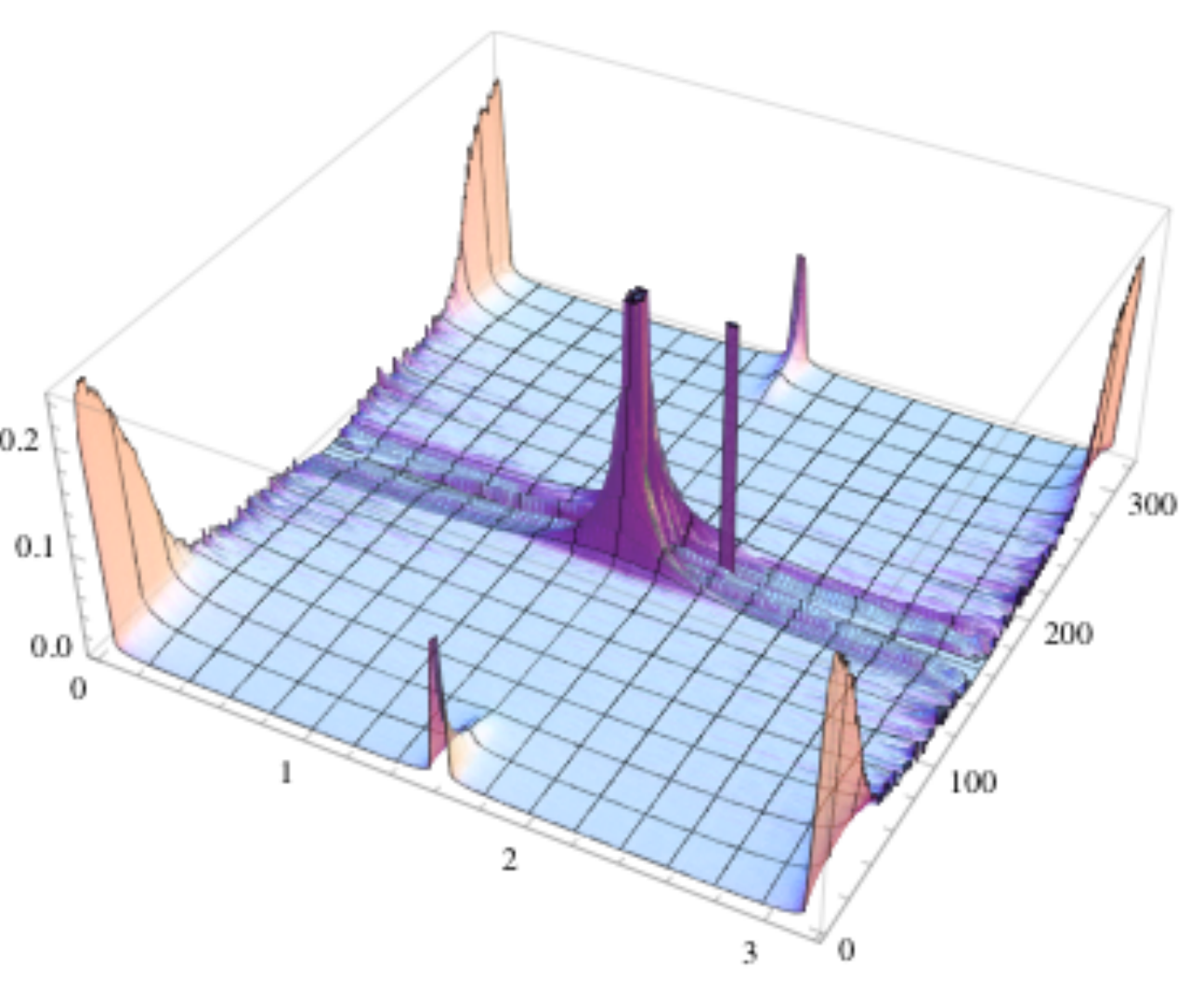} 
\caption{(color online) The probability distribution for the angle $f(\theta_{i,i+1})$ between nearest neighbour covariant Lyapunov vectors for a system of $80$ QOD hard disks at a density $\rho=0.003$ and temperature of $T=1$. The axis labeled $0$ to $3$ is the angle in radians, the axis labeled $0$ to $300$ is the vector number and the vertical axis is the probability.  The angle distributions initially show a strong preference for angles of $0$ and $\pi$ and a smaller preference for $\pi/2$. These preferences decay slowly with increasing vector number and disappear completely after vector number $50$.}
\label{N80_3}
\end{center}
\end{figure}

   In Fig. (\ref{N80_3}) we show the angle distributions between nearest neighbour vectors for a low density state in the kinetic regime for all pairs of vectors. 
   The probability distribution for the angle $\theta_{i,i+1}$ is at first glance uniform but with noticeable peaks near parallel ($0$, $\pi$) and orthogonal ($\pi /2$). 
   This system shows strong localization in the first $50$ or so Lyapunov vectors and the dominant first vector makes a major contribution to all the subsequent strongly localized vectors.
   This effect leads to the peaks for near parallel configurations of nearest neighbour pairs of vectors for $i=1,..,,50$,   becomes smaller with increasing $i$ and is absent after about $i=50$.
   The peaks near $0$ and $\pi$ suggests that pairs of vectors come arbitrarily close, almost parallel, in both the unstable manifold and the stable manifold.
   There is a similar central orthogonal peak at $\pi/2$ which decays very quickly with increasing $i$ and disappears completely after $i=25$ or $30$.
   While it is clear that a generic chaotic trajectory cannot develop an exact tangency, there appears to be a possible intermittency mechanism that inflates the measures near $0$ and $\pi$.
   In the centre of Fig. (\ref{N80_3}) between vectors $i=160$ to $i=200$ which correspond to the modes, we see a strong preference for close to orthogonal configurations but also a lone peak. 
   Apart from this region, Fig. (\ref{N80_3}) looks remarkably symmetrical with the distribution of $\theta_{i,i+1}$ almost to equal to the distribution for conjugate vector pair $\theta_{4N-i,4N-i+1}$.
   
\begin{figure}[htbp]
\begin{center}
	\includegraphics[scale=0.6] {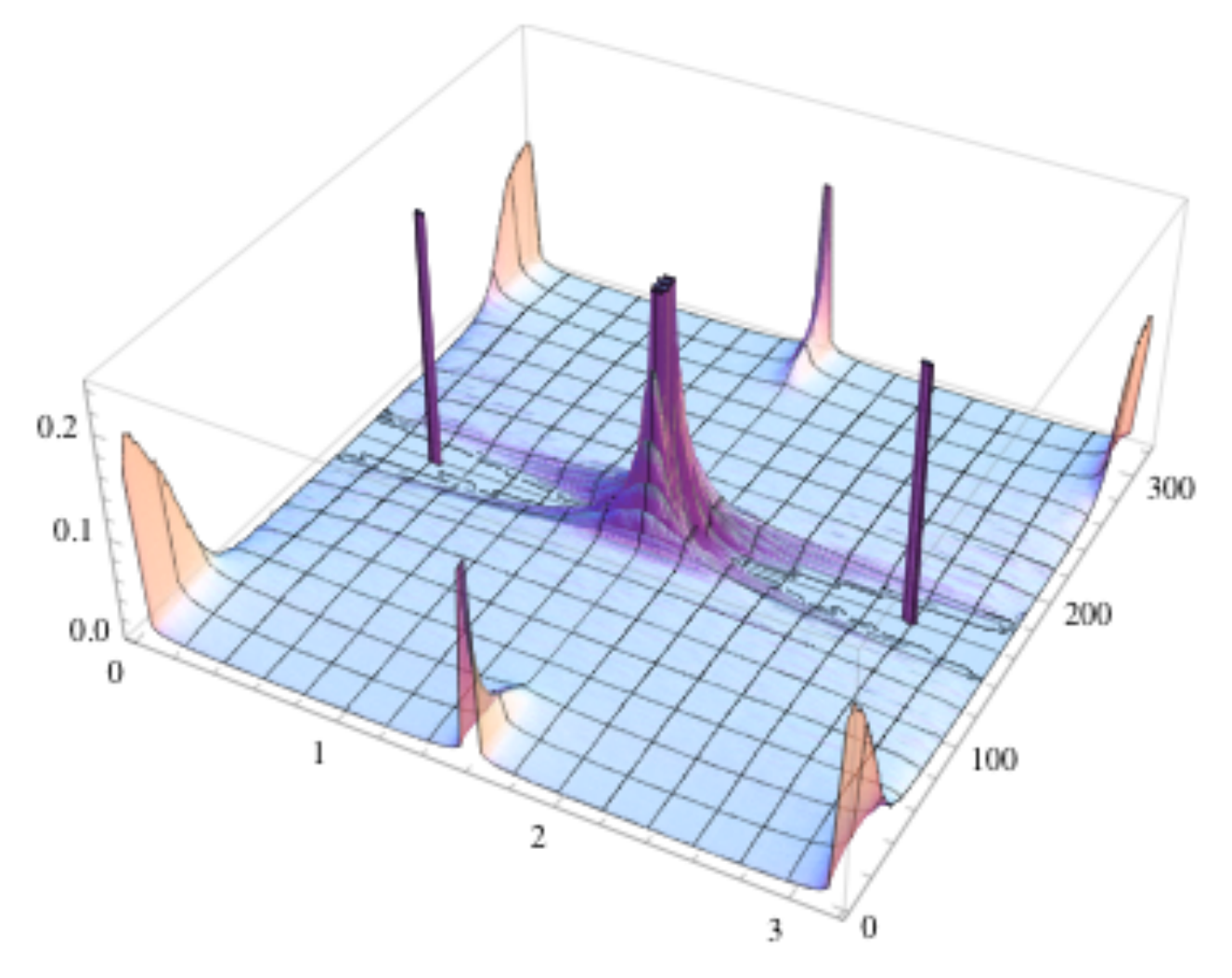} 
\caption{(color online) The probability distribution for the angle $\theta_{i,i+2}$ between second nearest neighbour covariant Lyapunov vectors for a system of $80$ QOD hard disks at a density $\rho=0.003$ and temperature of $T=1$.  The axes labels are the same as Fig. (\ref{N80_3}). Here the initial preference for $0$ and $\pi$ are less strong compared with the preference for $\pi/2$. Again these preferences decay and have disappeared completely after vector number $60$.}
\label{N80_4}
\end{center}
\end{figure}

  In Fig. (\ref{N80_4}) we show the angle distributions between second nearest neighbour vectors for a low density state in the kinetic regime for all of the possible second neighbour pairs. 
   Here we see similar effects in the distribution due to strong localization of the covariant Lyapunov vectors.
   The peaks in the first $50$ pairs of vectors at $0$ and $\pi$ are smaller than in Fig (\ref{N80_3}) while the peak at $\pi/2$ is a little larger and remains apparent for longer. 
   There is again evidence of strong structure in the mode region which is largely orthogonality between second neighbour vectors but there are two peaks with positions much closer to $0$ and $\pi$.
   Again there is symmetry between conjugate pairs of vectors angles with the distributions for $\theta_{i,i+1}$ and $\theta_{4N-i,4N-i+1}$ the same.

    As we move from 2nd neighbours to 3rd neighbours, 4th neighbours and beyond, the central peak at $\pi/2$ increases and the peaks at $0$ and $\pi$ decrease.

\begin{figure}[htbp]
\begin{center}
	\includegraphics[scale=0.2] {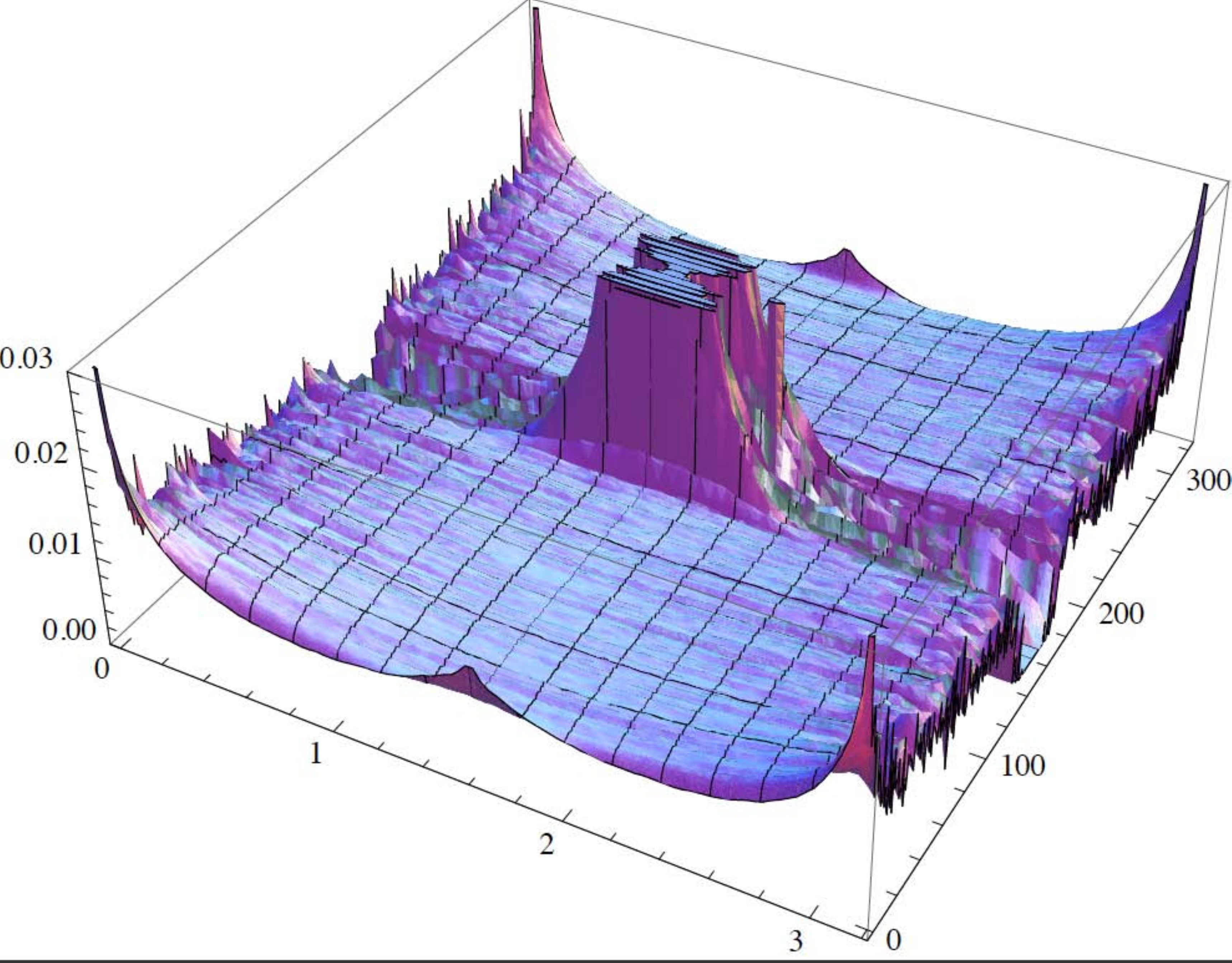} 
\caption{(color online) The probability distribution $f(\theta_{i,i+1})$ for the angle between nearest neighbour covariant Lyapunov vectors for a system of $80$ QOD hard disks at a density $\rho=0.8$ and temperature of $T=1$.  The axes labels are the same as Fig. (\ref{N80_3}). The three-dimensional plot of all vector pairs shows a high degree of symmetry between nearest neighbour angles corresponding to positive and negative pairs of vectors. Strong structure is also present in the angles between Lyapunov modes. Here we show results for pairs $i=1$ to $319$. At this density the angle distributions are only approximately constant, and the differences are associated localization. The distribution is slightly peaked at $0$, $\pi /2$ and $\pi$ but this feature disappears as the vector numbers increases from $1$ or decreases from $319$.}
\label{N80_1}
\end{center}
\end{figure}

\begin{figure}[htbp]
\begin{center}
	\includegraphics[scale=0.2] {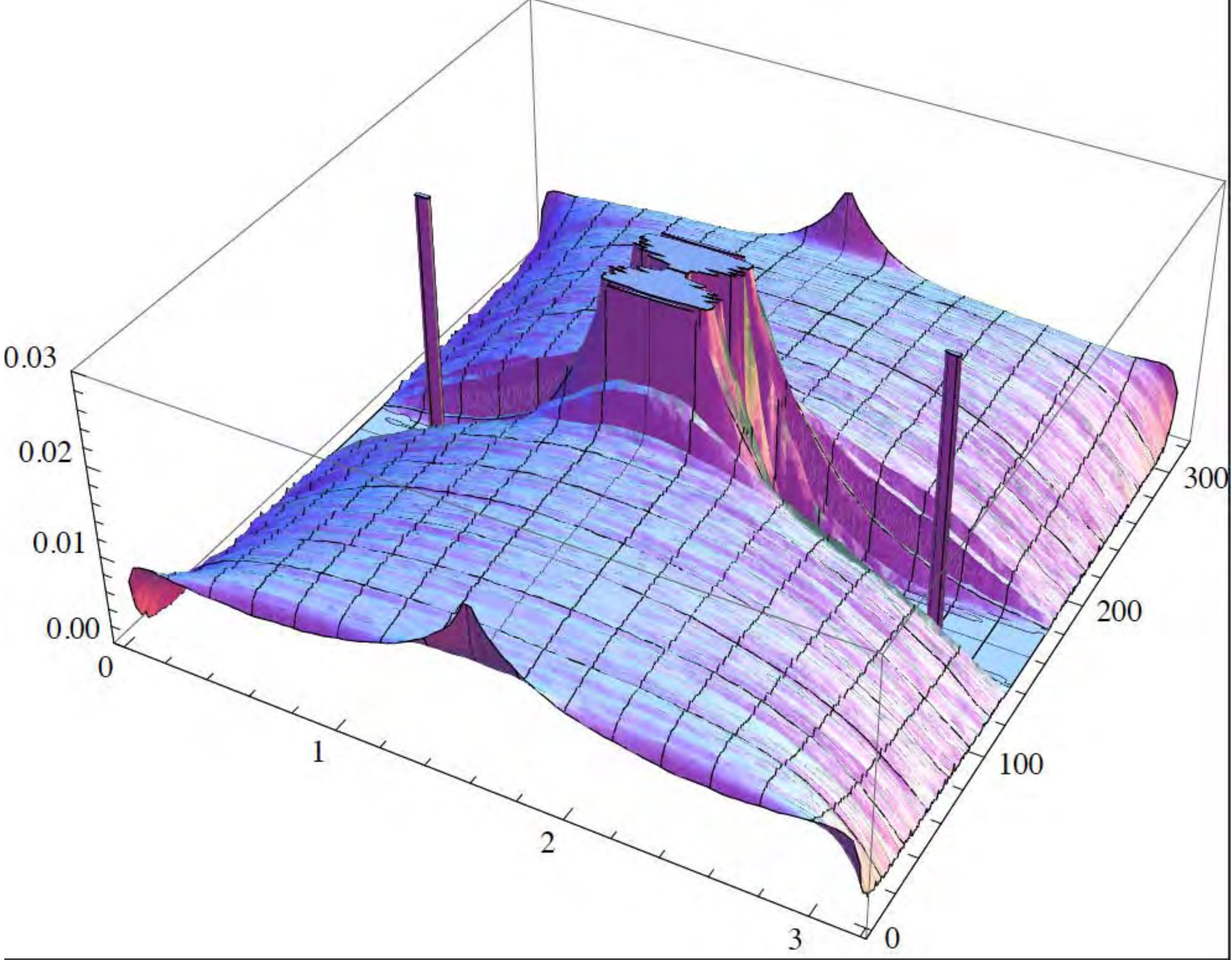} 
\caption{(color online) The probability distribution for the angle $\theta_{i,i+2}$ between second nearest neighbour covariant Lyapunov vectors for a system of $80$ QOD hard disks at a density $\rho=0.8$ and temperature of $T=1$.  The axes labels are the same as Fig. (\ref{N80_3}). At this density the distributions are very similar except for the first $10$ or so vectors. This is very similar to the systematic differences noted in Fig (\ref{N80_1}) in the first few vectors.}
\label{N80_2}
\end{center}
\end{figure}

  In  Fig. (\ref{N80_1}) we give a typical higher density result for the angle distribution of all $319$ nearest neighbour pairs of Lyapunov vectors $\theta_{i,i+1}$ at $\rho=0.8$.
   While the angle distributions are approximately constant in the continuous part of the spectrum there are features associated with remnants of the low density localization of individual vectors present at the beginning and end of the spectrum, and a preference for orthogonality for nearest neighbour vectors in the mode region.
   Otherwise the distributions show good symmetry for conjugate pairs of angles $\theta_{i,i+1}$ and $\theta_{4N-i,4N-i+1}$.
   In Fig. (\ref{N80_2}) we show the angle distributions $\theta_{i,i+2}$ between second nearest neighbour pairs of vectors in the high density regime for all pairs of second nearest neighbour vectors.
   Here the background distribution is approximately sinusoidal with deviations for a group of the first and last pairs of vectors, as well as strong structure in the mode region.
   The distributions show the same conjugate symmetry as before.

\begin{figure}[htbp]
\begin{center}
	\includegraphics[scale=0.5]{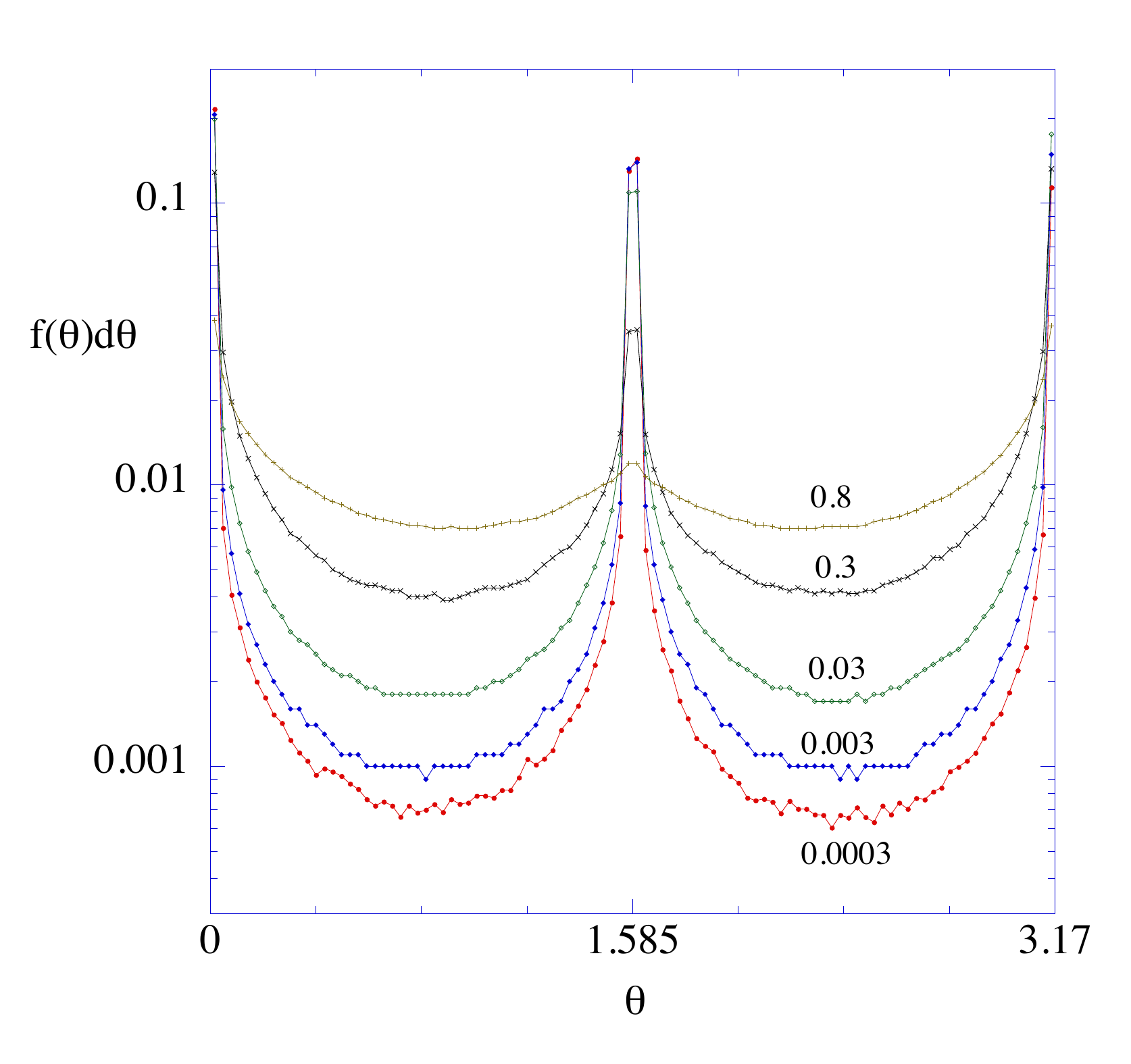}
\caption{(color online) The natural logarithm of the probability distribution for the angle $\theta_{1,2}$ for the first nearest neighbour pair of Lyapunov vectors for a system of $80$ hard disks at temperature $T=1$ as a function of density. At very low density, in the kinetic region, the distribution is sharply peaked at $0$, $\pi /2$ and $\pi$ but as the density increases the peaks become less sharp and smaller, until at $\rho = 0.8$ the distribution is almost constant in angle. Using a logarithmic scale on the vertical axis accentuates the differences between distributions at the lowest densities.}
\label{dist_1-2}
\end{center}
\end{figure}

   Fig. (\ref{dist_1-2}) shows how the distribution for the first nearest neighbour angle $\theta_{1,2}$ varies with density for an $80$ QOD particle system.
   At the lowest density of $0.0003$ the peaks at $0$, $\pi /2$ and $\pi$ dominate the distribution.
   Here the most localized configuration of a covariant vector is the dominant effect, so two vectors which are of most localized type, but supported on different pairs of particles are clearly orthogonal and contribute at $\pi/2$.
   If they are supported on the same pair of particles then this will lead to parallel or anti-parallel situations which contribute to peaks at $0$ and $\pi$.
   The parallel and anti-parallel configurations need to be more closely analysed as their nature cannot be assessed on the values of $\chi^{(j)}_{i}$ alone, but depend on the particle component contributions which lead to the values of $\chi^{(j)}_{i}$.
   Also different contributions will occur when the two vectors share only one particle in common.
   The distributions change systematically from low density, where the effects of localized covariant vectors dominates, to high density where the effects of localization have almost disappeared.

%
   

\section{Tangencies}

    So far the angle distributions we have considered are the angles between close neighbouring covariant vectors that are almost always contained in the same manifold, either both in the unstable manifold or both in the stable manifold.
    To consider the possibility of tangencies, which we interpret as the finite probability of an angle of $0$ or $\pi$ between one vector contained in the unstable manifold and one vector contained in the stable manifold, we consider the distribution of angles between conjugate covariant vectors $i$ and $4N-i+1$.
    This criteria has been proposed as a method to find the smallest angle between the stable and unstable spaces \cite{KK09} and probe the breakdown of hyperbolicity.
    
\begin{figure}[htbp]
\begin{center}
	\includegraphics[scale=0.6]{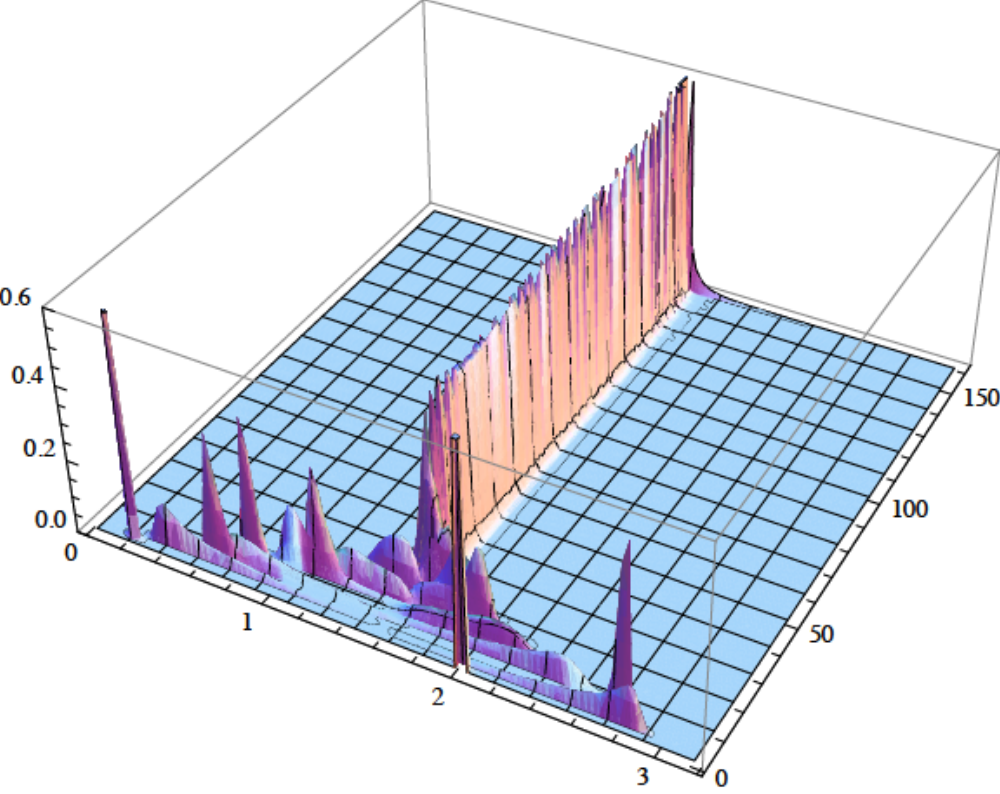}
\caption{(color online) The probability distribution for the angle $\theta_{i,4N-i+1}$ between conjugate pairs of vectors at a density of $\rho=0.8$ and temperature $T=1$ for a system of $N=80$ QOD hard disks. Here the axis labelled $1$ to $160$ is the vector number for the conjugate pair of modes. These begin at $1$ and $2$ with the zero modes and increase so that $160$ is the angle between the vectors corresponding to the largest and smallest exponents. In the mode region there are strong peaks but these are bounded away from $0$ and $\pi$. The continuous spectrum vectors all show the same orthogonality.  }
\label{con_ang_8}
\end{center}
\end{figure}

    In this QOD system we have seen that the smallest angles between stable and unstable covariant vectors occurs for conjugate pairs of vector modes, and typically the first conjugate pair of modes contains the smallest angle \cite{TM11b}.
   At high density the first mode is a conjugate pair of transverse (T) modes while at low density it is a conjugate pair of LP modes.
    The distribution of angles between all conjugate modes at high density is given in Fig. (\ref{con_ang_8}) and we see a clear distinction between the conjugate angle between modes and between vectors in the continuous region.
    There is a clear peak near zero for the first conjugate pair of T modes which is bounded away from zero and  each subsequent pair of modes gives a different peak, all of which are closer to $\pi/2$.
    The vectors that are not modes all show very good orthogonality, strongly peaked at $\pi/2$.

\begin{figure}[htbp]
\begin{center}
	\includegraphics[scale=0.6]{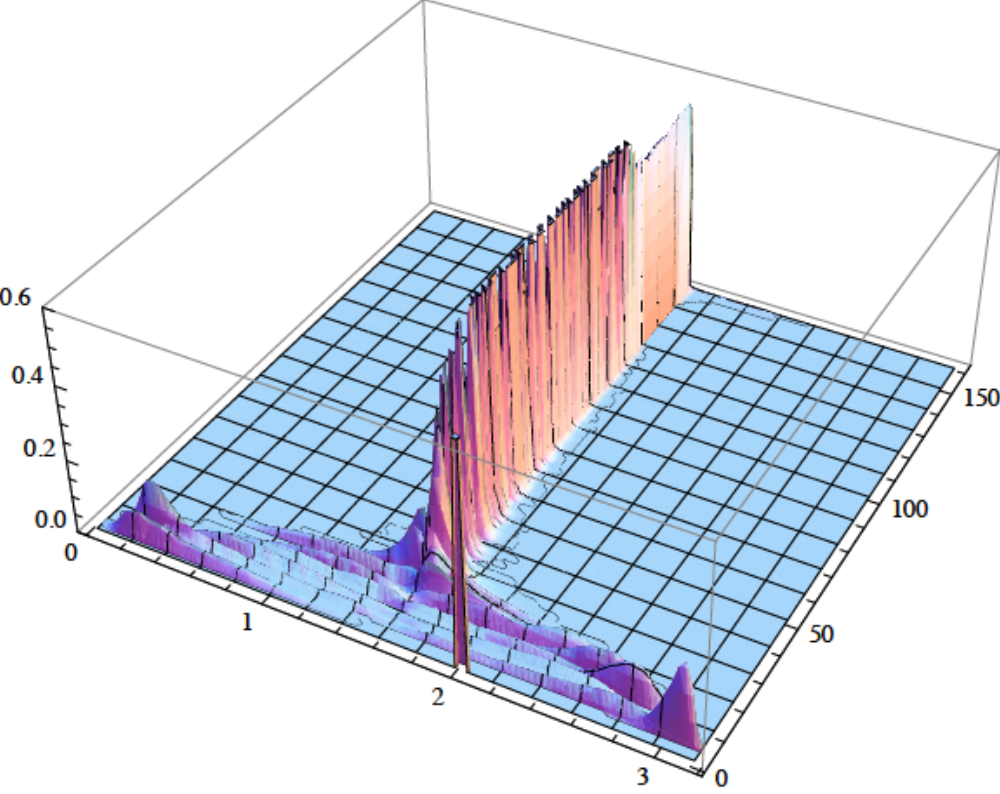}
\caption{(color online) The probability distribution for the angle $\theta_{i,4N-i+1}$ between conjugate pairs of vectors at a density of $\rho=0.003$ and temperature $T=1$ for a system of $N=80$ hard disks. This shows similar systematic changes with mode number compared with the high density result. There is a feature in the angle distribution associated with the strongly localized vectors in the positive and negative halves of the spectrum.}
\label{con_ang_003}
\end{center}
\end{figure}

   At lower density, $\rho=0.003$, see Fig. (\ref{con_ang_003}), the distribution between the first conjugate pair of modes (here LP modes) approaches $0$ and $\pi$ but the distribution goes to zero at both ends, so again no possibility of tangency although the distribution is no longer strongly bounded away from $\pi$.
   Indeed the distributions for all pairs of are much broader and less peaked than at high density.
   However, here there is a clear gap and no tangencies are possible.
   The vectors in the continuous part of the spectrum all show strong orthogonality with dominant peaks at $\pi/2$.

\begin{figure}[htbp]
\begin{center}
	\includegraphics[scale=0.6]{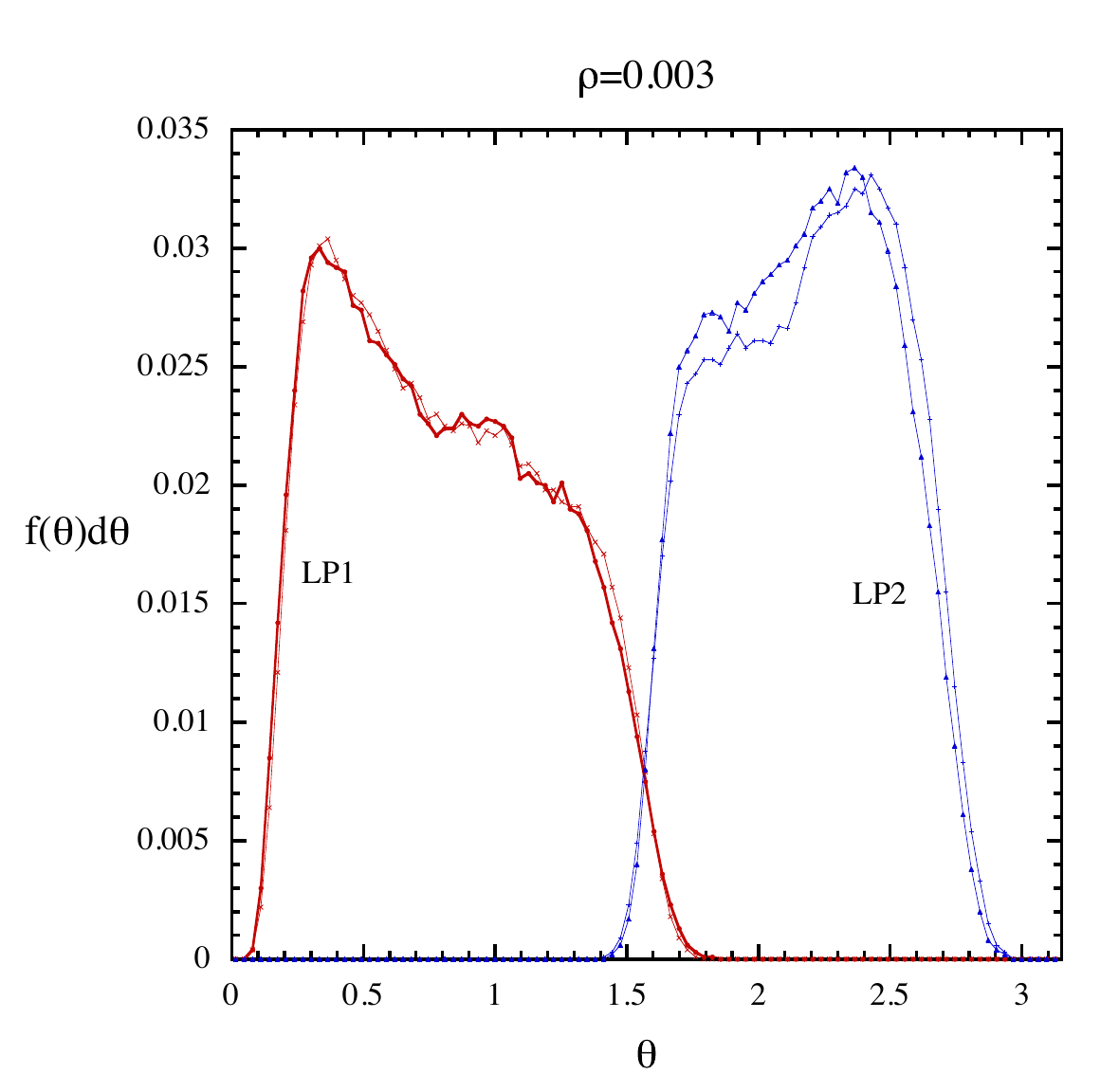}
\caption{(color online) The probability distribution for the angle $\theta$ between first two conjugate pairs of LP modes at a temperature $1$ and density of $\rho=0.003$, for a system of $80$ hard disks. Notice that the distribution is bounded away from $0$ (or $\pi$) and goes to zero fast enough to avoid tangencies.}
\label{LPmode}
\end{center}
\end{figure}

   The angle distribution between the first conjugate LP modes at lower density ($\rho=0.003$) are shown in Fig. (\ref{LPmode}). 
   Although these distributions are the $3^{rd}$ line in Fig. (\ref{con_ang_003}), we have enlarged the vertical scale so that it is clear that the distributions are bounded away from possible tangencies at $0$ and $\pi$, despite the fact that the distributions are broad and largest close to these limits.

   Previous theoretical studies of the time evolution of a pair of conjugate modes \cite{CTM10} showed that the dynamics approximately separates from the dynamics of the whole tangent space, and can be developed for a pair of T modes by considering the product of $2\times 2$ reduced $R$ matrices that produce the backward time covariant evolution \cite{TM11b}.
   These matrices have the form
\begin{equation}\label{Rm}
R_{m}^{-1} = \left( \begin{array}{cc} -\zeta_{m}^{-1} &  c_{m} \zeta_{m} \\0  & \zeta_{m} \end{array} \right).
\end{equation}
where $\zeta_{m}$ is the local expansion rate for the $m^{th}$ cycle of free-flight and collision, and $c_{m}$ is related to the free flight time (usually a little smaller than $\tau_{m}$).
   The cosine of the angle between these conjugate modes can be written as
\begin{equation}\label{thetam}
   \cos \theta_{i,j} = \frac {f} {\sqrt{f^{2} + 1}}
\end{equation}
where the function $f$ depends on the sets of values $\{ \zeta_{j} \}$ and $\{ c_{j} \}$ contained in all the $R_{j}$  matrices along the path.
   For an evolution of $n-m$ steps backward from $m$ the function is
\begin{equation}\label{fmn}
   f(m,n, \{\zeta_{j} \},\{ c_{j} \} ) = c_{n+1} +  \sum_{j=n+2}^{m} \zeta_{n+1}^{2}....\zeta_{j-1}^{2} c_{j}.
\end{equation}
   In the limit as $n \rightarrow \infty$ the vector converges to the covariant vector 
and as long as $\zeta_{j} > 1$, the angle $\theta_{i,j}$ is bounded away from $0$ or $\pi$.
   To see the boundedness explicitly as $n \rightarrow \infty$ we set $\zeta_{j} = \zeta$ and $c_{j} = \tau$ for all $j$ and then $f = \tau/(1-\zeta^{-2})$ which is finite as long as $\zeta \neq 1$.
   As long as $f$ remains finite in Eqn. (\ref{thetam}), then cosine of the angle is bounded away from $1$ or $-1$.
   

\section{Conclusions}

   We have studied the Lyapunov spectrum and vectors, both GS and covariant, for QOD systems over the full range of physical densities.
   The first observation is that the exponents in the step region depend linearly on the mode number for both T and LP modes and given the slopes of these linear dependences as a function of density, we can predict the order in which the modes appear, or equivalently the order in which the steps appear in the spectrum. 
   Next we observe that both the Lyapunov spectrum and the localization obey simple scaling relations in the kinetic regime for the same group of vector numbers - typically numbers 40 to 160 for an 80 particle system.
   Here the Lyapunov spectrum is linear in density and the localization is independent of density.

   The average localizations for covariant modes is generally significantly smaller and thus more localized than the results for  the GS modes.
    This is observed at low density in the kinetic region and also at high density.
      
   The strongly localized configurations for GS vectors consisting of a dominant contribution from two neighbouring particles are again apparent and the covariant vectors being linear combinations of GS vectors show the same feature. 
   However, the removal of the orthogonality condition in going from GS to covariant vectors allows one strongly localized configuration to dominate a whole group of covariant vectors whereas it can only contribute to one GS vector. 
   This is evident for the low density average localization where many of the strongly localized vectors have essentially the same average localization, and is connected with the instantaneous localization.
   
   There is very clear evidence of asymptotic strong localization in the kinetic regime for the first $20\%$ vectors of the covariant spectrum where the participation ratio $N W(N)$ is independent of system size.
   
    The simple scheme for determining the angle distributions between neighbouring covariant vectors proposed here ($ \rho (\theta _d ) = \sin^{d-1} (\theta_d)$) is useful in understanding the underlying behaviour of the distributions but the results are affected by the degree of localization and this is determined by the density.
    At low density, strong localization has a significant effect however, there is some strong localization effect at all densities - limited to the first and last few vectors at high density.
    
    There is no evidence for tangencies in the QOD system.
    Indeed, if two vectors become parallel then the matrix $V_{m}$ loses rank and the method will fail.
    This is never observed for this QOD system in finite time. 
    The closest approach to tangency is for the vectors associated with the first positive and negative modes.
    Only at low densities in the kinetic regime do the first vectors approach $0$ or $\pi$.
    The angles between conjugate vectors in the unstable manifold and the stable manifold show a high degree of orthogonality suggesting that two manifolds are generally orthogonal.

%
%

\begin{acknowledgments}

The author thanks F. Ginelli, K. A. Takeuchi and H. Chat\'e for helpful discussions and Service de Physique de l'\'Etat Condens\'e for partial financial support.
\end{acknowledgments}

%
%

%

\begin{thebibliography}{99} 


\bibitem{Ben} G. Benettin, L. Galgani, and J. -M. Strelcyn, Phys. Rev. A \textbf{14}, 2338 (1976); G. Benettin, L. Galgani, A. Giorgilli and J. -M. Strelcyn, C. R. Acad. Sci. Ser. A \textbf{286},  431 (1978); G. Benettin, L. Galgani, A. Giorgilli and J. -M. Strelcyn, Meccanica \textbf{15}, 9 (1980); \& \textbf{15}, 21 (1980). 

\bibitem{Shi79} I. Shimada and T. Nagashima, Prog. Theor. Phys. \textbf{61}, 1605 (1979). 



\bibitem{Posg} H. A. Posch and R. Hirschl, in \textit{Hard ball systems and the Lorentz gas}, edited by D. Sz\'asz (Springer, Berlin, 2000), p. 279;  Lj. Milanovi\'c and H. A. Posch, J. Mol. Liquids, \textbf{96-97}, 221 (2002); C. Forster, R. Hirschl, and H. A. Posch, ÒAnalysis of Lyapunov modes for hard-disk fluidsÓ  Proceedings for the XIV International Congress on Mathematical Physics, ICMP 2003, edited by J-C. Zambrini (University of Lisbon, Portugal) page 423.

\bibitem{For04} C. Forster and H. A. Posch, New J. Phys. \textbf{7}, 32 (2005).

\bibitem{TMg}  T. Taniguchi and G. P. Morriss, Phys. Rev. E \textbf{68}, 026218 (2003); Phys. Rev. E \textbf{71},  016218 (2005); Phys. Rev. Lett., \textbf{94}, 154101 (2005) ; European Physical Journal B, \textbf{50},  305 (2006).

\bibitem{MT09} D. J. Robinson and G. P. Morriss, J. Stat. Phys., \textbf{131}, 1. (2008); G. P. Morriss and D. Truant, J. Stat. Mech., (2009), P02029.

\bibitem{CTM10} T. Chung, D. Truant and G. P. Morriss, Phys. Rev. E \textbf{81}, 066208, (2010); T. Chung, D. Truant and G. P. Morriss, Phys. Rev. E \textbf{83}, 046216, (2011).

\bibitem{MT11a} G. P. Morriss and D. Truant, Molecular Simulations \textbf{37}, 277 (2011).


\bibitem{TM11b} D. Truant and G. P. Morriss, J. Stat. Mech. P01014 (2011).



\bibitem{TMth} T. Taniguchi, C. P. Dettmann, and G. P. Morriss, J. Stat. Phys. \textbf{109}, 747 (2002); T. Taniguchi and G. P. Morriss,  Phys. Rev. E \textbf{65}, 056202 (2002); Phys. Rev. E \textbf{66}, 066203 (2002)


\bibitem{Eck05} J. -P. Eckmann, C. Forster, H. A. Posch, and E. Zabey, J. Stat. Phys. \textbf{118}, 813 (2005). 


\bibitem{Yan05} H. -l. Yang and G. Radons, Phys. Rev. E \textbf{71}, 036211 (2005).




\bibitem{ECM90} D. J. Evans, E. G. D. Cohen, and G. P. Morriss, Phys. Rev. A \textbf{42}, 5990 (1990). 

\bibitem{ECM93} D. J. Evans, E. G. D. Cohen, and G. P. Morriss, Phys. Rev. Lett. \textbf{71}, 2401 (1993). 

\bibitem{DM96} C. P. Dettmann and G. P. Morriss, Phys. Rev. E \textbf{53}, R5545 (1996). 

\bibitem{GC95} G. Gallavotti and E. G. D. Cohen, J. Stat. Phys. \textbf{80}, 931 (1995).
   

\bibitem{AM78} R. Abraham and J. Marsden, {\it Foundations of mechanics} (Reading, Mass: Benjamin/Cummings, 1978).


\bibitem{Kry79} N. S. Krylov, {\it Works on the Foundations of Statistical Physics}, Translated by A. B. Migdal, Ya. G. Sinai, and Yu. L. Zeeman, Princeton Series in Physics (Princeton University Press, Princeton, New Jersey, 1979).


\bibitem{Man85} P. Manneville, Lecture Notes in Physics, \textbf{230}, 319 (1985).    

\bibitem{Kan86} K. Kaneko, Physica D \textbf{23}, 436 (1986).    




\bibitem{Cha93} H. Chat\'{e}, Europhys. Lett. \textbf{21}, 419 (1993).




\bibitem{TM03b} T. Taniguchi and G. P. Morriss,  Phys. Rev. E \textbf{68}, 046203 (2003); 


\bibitem{TM06} T. Taniguchi and G. P. Morriss,  Phys. Rev. E \textbf{73}, 036208 (2006); Physica A \textbf{375} (2007) 563,


\bibitem{WS07} C. L. Wolfe and R. M. Samelson, Tellus, \textbf{59A}  355 (2007).

\bibitem{GPTCLP07} F. Ginelli, P. Poggi, A. Turchi, H. Chat\'e, R.Livi and A. Politi, Phys. Rev. Lett. \textbf{99}, 130601 (2007).   

\bibitem{Sz07} I. G. Szendro, D. Pazo, M. A. Rodriguez and J. M. Lopez, Phys. Rev. E \textbf{76} 025202 (2007); D. Pazo,  I. G. Szendro, J. M. Lopez and M. A. Rodriguez, Phys. Rev. E \textbf{78} 016209 (2008).


\bibitem{YT09} H. -l. Yang, K. A. Takeuchi, F. Ginelli, H. Chat\'e and G. Radons, Phys. Rev. Lett. \textbf{102} 074102 (2009) .

\bibitem{TGC09} K. A. Takeuchi, F. Ginelli, and H. Chat\'e, Phys. Rev. Lett. \textbf{103}, 154103 (2009).

\bibitem{KK09} P. V. Kuptsov and S. P. Kuznetsov, Phys. Rev. E \textbf{80}, 016205 (2009).


\bibitem{BP10} H. Bosetti and H. A. Posch, Chem. Phys. \textbf{375}, 296 (2010).

\bibitem{TYGRC11} K. A. Takeuchi, H. -l. Yang, F. Ginelli, G. Radons and H. Chat\'e, (arXiv:1107.2567).

\bibitem{TCGPT11} K. A. Takeuchi, H. Chat\'e, F. Ginelli, A. Politi and A. Torcini, Phys. Rev. Lett. \textbf{107}, 124101 (2011).



\bibitem{Gas98} P. Gaspard,  {\it Chaos, scattering and statistical mechanics}  (Cambridge University press, 1998).

\bibitem{Dor99} J. R. Dorfman,  {\it An introduction to chaos in nonequilibrium statistical mechanics}  
   (Cambridge University press, Cambridge, 1999).


\bibitem{EM08} D. J. Evans and G. P. Morriss, {\it Statistical mechanics of nonequilibrium liquids}  2nd Ed. 
   (Cambridge University press, Cambridge, 2008).


\bibitem{Del96} Ch. Dellago, H. A. Posch, and W. G. Hoover, Phys. Rev. E \textbf{53}, 1485 (1996); Ch. Dellago and H. A. Posch,  Physica A \textbf{240}, 68 (1997). 


\bibitem{vB97} H. van Beijeren, J. R. Dorfman, H. A. Posch and Ch. Dellago, Phys. Rev. E \textbf{56} 5272 (1997);  J. R. Dorfman, A. Latz and  H. van Beijeren, CHAOS \textbf{8} 444 (1998): H. van Beijeren and J. R. Dorfman, J. Stat. Phys. \textbf{108} 767 (2002); A. S. de Wijn, Phys. Rev. E \textbf{71} 046211 (2005).

\bibitem{FMP04} Ch. Forster, D. Mukamel and H. A. Posch, Phys. Rev. E \textbf{69}, 066124 (2004)


\bibitem{TC10} G. Radons, K. A. Takeuchi and H. Chat\'{e}, unpublished.

\end{thebibliography}
\end{document}